\documentclass[apj,iop]{emulateapj}

\usepackage{graphicx} 
\usepackage{apjfonts} 
\usepackage{hyperref}
\usepackage{subfigure}

\usepackage{xcolor}

\shorttitle{Emergence of internetwork magnetic fields through the solar atmosphere} 

\shortauthors{Go\v{s}i\'{c} et al.}

\begin{document}

\title{Emergence of internetwork magnetic fields through the solar atmosphere}

\author{M.~Go\v{s}i\'{c}}
\affil{Lockheed Martin Solar and Astrophysics Laboratory, Palo Alto, CA 94304, USA; gosic@baeri.org}
\affil{Bay Area Environmental Research Institute, Moffett Field, CA 94035, USA}

\author{B.~De Pontieu}
\affil{Lockheed Martin Solar and Astrophysics Laboratory, Palo Alto, CA 94304, USA}
\affil{Institute of Theoretical Astrophysics, University of Oslo, P.O. Box 1029 Blindern, NO-0315 Oslo, Norway}
\affil{Rosseland Centre for Solar Physics, University of Oslo, P.O. Box 1029 Blindern, NO-0315 Oslo, Norway}

\author{L.~R.~Bellot Rubio}
\affil{Instituto de Astrof\'{\i}sica de Andaluc\'{\i}a (IAA-CSIC), Apdo.\ 3004, 18080 Granada, Spain} 

\author{A.~Sainz Dalda}
\affil{Lockheed Martin Solar and Astrophysics Laboratory, Palo Alto, CA 94304, USA}
\affil{Bay Area Environmental Research Institute, Moffett Field, CA 94035, USA}

\author{S.~Esteban Pozuelo}
\affil{Instituto de Astrof\'{\i}sica de Andaluc\'{\i}a (IAA-CSIC), Apdo.\ 3004, 18080 Granada, Spain} 
\affil{Instituto de Astrof\'{\i}sica de Canarias, C/ V\'ia L\'actea s/n, 38205, La Laguna, Tenerife, Spain}
\affil{Departamento de Astrof\'{\i}sica, Universidad de La Laguna, 38205, La Laguna, Tenerife, Spain}

\begin{abstract}
Internetwork (IN) magnetic fields are highly dynamic, short-lived magnetic structures that populate the interior of supergranular cells. Since they emerge all over the Sun, these small-scale fields bring a substantial amount of flux, and therefore energy, to the solar surface. Because of this, IN fields are crucial for understanding the quiet Sun (QS) magnetism. However, they are weak and produce very small polarization signals, which is the reason why their properties and impact on the energetics and dynamics of the solar atmosphere are poorly known. Here we use coordinated, high-resolution, multiwavelength observations obtained with the Swedish 1-m Solar Telescope (SST) and the \textit{Interface Region Imaging Spectrograph} (IRIS) to follow the evolution of IN magnetic loops as they emerge into the photosphere and reach the chromosphere and transition region. We studied in this paper three flux emergence events having total unsigned magnetic fluxes of $1.9\times10^{18}$, $2.5\times10^{18}$, and $5.3\times10^{18}$~Mx. The footpoints of the emerging IN bipoles are clearly seen to appear in the photosphere and to rise up through the solar atmosphere, as observed in \ion{Fe}{1} 6173 \AA\/ and \ion{Mg}{1} b$_2$ 5173 \AA\/ magnetograms, respectively. For the first time, our polarimetric measurements taken in the chromospheric \ion{Ca}{2} 8542 \AA\/ line provide direct observational evidence that IN fields are capable of reaching the chromosphere. Moreover, using IRIS data, we study the effects of these weak fields on the heating of the chromosphere and transition region.
\end{abstract}


\section{Introduction}

During the 1970s, the solar community made decisive steps toward a better understanding of solar magnetism. The pioneering works of \cite{LivingstonHarvey1971}, \cite{LivingstonHarvey}, and \cite{Smithson1975} clearly showed the presence of small-scale magnetic features in the QS. They called them \textit{inner-network} or \textit{internetwork} (IN) fields. This new component of the solar magnetic field was described as an ensemble of mixed-polarity magnetic elements, independent of the surrounding network (NW) patches, but with a tendency to move toward them. Despite the low spatial and temporal resolution of the observations, these results provided insight into the temporal evolution and physical properties of IN fields. They still stand quite accurate, although they have been extended in many ways (see \citealt{BellotRubioOrozcoSuarez2019} for a review).

IN magnetic features with fluxes in the range from $10^{16}$~Mx to $10^{18}$~Mx regularly appear on the solar surface in bipolar form, i.e., in the form of magnetic loops \citep{Litesetal1996, Centenoetal2007, MartinezGonzalezBellotRubio2009, MartinezGonzalezetal2010, Gomoryetal2010, Guglielminoetal2012, Fischeretal2019}. The observational signature of the loops is a linear polarization signal above or at the edges of a granular cell, followed by two circular polarization (CP) patches on its flanks. The former corresponds to the loop top, while the latter are associated with the loop footpoints. As the top of the loop ascends through the solar photosphere, the linear polarization patch disappears. In the meantime, the footpoints are dragged by convective motions toward intergranular lanes, separating from each other following a more or less straight trajectory. These loops may also emerge as a group of mixed-polarity patches within a relatively small region (see e.g. \citealt{Wang1988}, \citealt{HarveyMartin1973}, \citealt{Harveyetal1975}, \citealt{Title2000}, \citealt{Hagenaar2001}, \citealt{Gosic2015}), forming clusters of IN magnetic elements with typical total fluxes between $10^{18}$ and $10^{19}$~Mx. 

Since IN bipolar structures are spread all over the Sun, they may hold a significant fraction of the total magnetic energy stored in the photosphere. As discussed by \cite{TrujilloBueno2004}, the IN energy budget is expected to be sufficient to globally balance the radiative losses in the upper QS atmosphere. Based on potential field extrapolations, one can expect small-scale IN loops to make it up to heights of 2-3~Mm above the continuum forming layer \citep{MartinezGonzalezetal2010}. The same result is obtained from magnetofrictional simulations (Go\v{s}i\'{c} et al., in prep.). Indeed, from Hinode/SP observations, \cite{MartinezGonzalezBellotRubio2009} estimated that about 23\% of the detected loops (from $10^{16}$ to $10^{18}$~Mx) reach the height of formation of the \ion{Mg}{1} b$_2$ 5173 \AA\/ line (upper photosphere/lower chromosphere). These authors did not have access to polarization signals in the chromosphere and above to further follow those loops. Instead, the loop footpoints were detected indirectly in \ion{Ca}{2} H line-core intensity maps as bright points. Similar emission features associated with small-scale flux emergence have been seen recently in other lines \citep{Kontogiannisetal2019}, although at lower resolution than that provided by instruments such as the Swedish 1-m Solar Telescope \citep[SST;][]{Scharmeretal2003}. However, the presence of polarization signals due to IN magnetic fields in the chromosphere is still not observationally confirmed due to insufficient sensitivity of the available measurements. It also remains unknown if these fields can rise up to the mid chromosphere, transition region and corona. Observations obtained in active regions have revealed that granular-sized loop structures with total flux on the order of $10^{18}$~Mx may reach the chromosphere \citep{Wang2020}, while the Interface Region Imaging Spectrograph \citep[IRIS;][]{DePontieuetal2014} in coordination with ground-based telescopes have shown that such loops may reach even the transition region and locally heat the corona \citep{Ortizetal2014, VargasDominguez2014, delaCruzRodriguezetal2015a, Ortizetal2016}. Something similar may happen in IN regions.

A number of theoretical studies have discussed small-scale magnetic flux emergence from the convection zone through the solar atmosphere \citep{SteinNordlund2006, Cheungetal2007, MartinezSykoraetal2008, MartinezSykoraetal2009, Isobeetal2008, TortosaAndreuMorenoInsertis2009}. Yet, MHD simulations do not consistently show IN magnetic loops reaching the chromosphere, let alone the transition region or corona. Recent 3D numerical models of QS magnetoconvection indicate that it is very difficult for IN fields to ascend to the chromosphere due to the lack of magnetic buoyancy. \cite{MorenoInsertisetal2018} detected only one such case (see also \citealt{MartinezSykoraetal2019}). In general, small-scale IN fields lose spatial coherence after rising a few hundred km above the photosphere. In the simulations presented by \cite{SteinNordlund2006}, magnetic loops disintegrate while passing through the photosphere and never reach the upper solar atmosphere. Other MHD models, such as those published by \cite{Isobeetal2008}, not only show that IN magnetic loops reach chromospheric levels, but also suggest that IN loops can heat the chromospheric plasma through reconnection with the ambient magnetic fields. These findings are supported by \cite{Amarietal2015}. In their model, IN magnetic fields emerge as a result of local dynamo action, and contribute to the coronal heating thanks to numerous small-scale eruptions driven by magnetic reconnection. Further studies are needed to resolve the discrepancies between these models, and establish which one is the most accurate regarding the presence and nature of IN fields in the photosphere, including their impact on the solar atmosphere.

In this paper we present the temporal evolution of three bipolar IN magnetic structures. To determine how far these fields can climb through the solar atmosphere, we employ simultaneous, multi-instrument, multi-wavelength observations. This is achieved using SST and IRIS. With these instruments we have studied the spatio-temporal evolution of the weak, granular-sized IN fields at high spatial, spectral, and temporal resolution, with an unprecedented sensitivity, sufficient to detect for the first time the polarization signals of IN magnetic fields in the chromosphere. Thanks to these coordinated observations, we can observe the solar atmosphere from the photosphere up to the transition region. Thus, they have allowed us to detect and quantify for the first time the chromospheric heating associated with newly emerging IN magnetic loops. 

The observational data used in this paper are described in Section \ref{sect2}. In Section \ref{sect3} we explain how we identify, classify and track magnetic bipoles. Section \ref{sect4} gives examples of newly emerged IN loops, and show how they ascend to the chromosphere and transition region. We also provide observational evidence for heating resulting from emerging flux and quantify it. Finally, we present our conclusions in Section \ref{sect5}.

\section{Observations and data processing}
\label{sect2}

In this work we employ observations obtained with SST and IRIS on 2014 May 16 that continuously monitored a large area of the QS at disk center for $2.5$~hr. The data sequences covered several supergranular cells, one of which is completely visible in the field of view (FOV) of about $60\arcsec \times 60\arcsec$. The data sets reveal the spatio-temporal evolution of QS patches from the photospheric layers up to the upper chromosphere. We present these observations below and refer the reader to \cite{Gosicetal2018} for more details. 

\subsection{SST observations}

We obtained spectropolarimetric measurements using the CRisp Imaging SpectroPolarimeter \citep[CRISP;][]{Lofdahl2002, Scharmeretal2008}, mounted on the SST. CRISP takes high spatial-resolution monochromatic images within a narrow passband in the spectral range between 5000 to 8600 \AA\/, with a pixel size of $0\farcs057$ at 6300 \AA\/. Our observational sequence was recorded between 07:23:41 and 10:28:44 UT.

CRISP measured the full Stokes profiles of the \ion{Fe}{1} 6173 \AA\/, \ion{Mg}{1} b$_2$ 5173 \AA\/, and \ion{Ca}{2} 8542 \AA\/ lines in a QS region at the disk center. In the H$\alpha$ 6563 \AA\/ line, CRISP operated in intensity-only mode. The \ion{Fe}{1} 6173 line was sampled at 11 wavelength positions in steps of 28 m\AA\/, plus a continuum point at $+532$ m\AA\/, to monitor the lower photosphere. This accounts for a scanning time of 20~s. The upper photosphere was observed in the \ion{Mg}{1} b$_2$ line at 10 wavelength positions, which required 11~s to complete a scan. We selected two points at $\pm 50$~m\AA\/ from the line core, seven in the wings with steps of 100 m\AA\/, and one wavelength position at $-700$ m\AA\/. The chromospheric measurements in the \ion{Ca}{2} 8542 line were obtained within 19~s. The Stokes parameters were sampled at 17 wavelength positions in steps of 100 m\AA\/ plus a continuum point at $+2.4$ \AA\/. Intensity maps in H$\alpha$ were recorded with a scanning time of 5~s using 21 spectral positions in steps of 100 m\AA\/. The scanning times add up to a cadence of 55~s, resulting in 211 cycles in total. It is important to highlight that we used very long integration times for the \ion{Ca}{2} 8542 \AA\/ line to detect weak QS magnetic fields in the chromosphere. We used the polarimetric signals in the continuum to estimate the noise level ($\sigma$) as the standard deviation of Stokes $Q$, $U$ and $V$ normalized to the continuum intensity. The final noise levels achieved in the \ion{Ca}{2} 8542 \AA\/ line are $\sigma_{Q}=1.2\times10^{-3}$, $\sigma_{U}=1.4\times10^{-3}$ and $\sigma_{V}=1.4\times10^{-3}$, respectively.

The data were reduced using the CRISPRED pipeline \citep{delaCruzRodriguezetal2015b}, including corrections by \cite{vanNoortRouppevanderVoort2008} and \cite{Shineetal1994}. The images were restored applying the Multi-Object, Multi-Frame Blind-Deconvolution technique \citep{vanNoortetal2005}. We also corrected residual seeing motions \citep{Henriques2012}, removed instrumental polarization employing the telescope polarization model \citep{delaCruzRodriguezetal2015b, Selbing2005}, and corrected residual crosstalk from $I$ to $Q$, $U$, and $V$ using the Stokes parameters in the continuum. Finally, we removed five and three-minute oscillations from the photospheric and chromospheric observations, respectively, using a subsonic filter \citep{1989ApJ...336..475T, 1992AA...256..652S}.

Magnetograms $M$ and Dopplergrams $D$ were constructed from Stokes $I$ and $V$ filtergrams using the classical approach:

\begin{equation}
\label{mag_eq}
M=\frac{1}{2}\left(
\frac{V_{\text{blue}}}{I_{\text{blue}}}-\frac{V_{\text{red}}}{I_{\text{red}}}
\right),
\label{eq1}
\end{equation}

\begin{equation} 
D = \frac{I_{\rm blue}-I_{\rm red}}{I_{\rm blue}+I_{\rm red}}.
\end{equation} 

\noindent where ``blue'' stands for the measurements obtained in the blue wing of the line and ``red'' for those in the red wing. Magnetograms in the \ion{Fe}{1} 6173 \AA\/ line were derived using the measurements taken at $\pm84$~m\AA\/. In the case of the \ion{Mg}{1} b$_2$ and \ion{Ca}{2}~8542~\AA\/ magnetograms, we averaged over two spectral positions in the wings of the lines ($\pm100$ and $\pm200$~m\AA\/). To reduce the noise, the magnetograms were smoothed using a $3\times3$ Gaussian-type spatial kernel.

Maps of linear polarization signal $\rm LP$ are created as ${\rm LP}= \sum_{i=1}^{4} \sqrt{Q(\lambda_i)^2 + U(\lambda_i)^2}/I(\lambda_i)/4$, taking into account the first four wavelengths around the core of the \ion{Fe}{1} and \ion{Mg}{1}~b$_2$ lines, i.e., $(\pm56, \pm140)$ and $(\pm50, \pm300)$, respectively.

We calculate circular polarization maps as ${\rm CP}=\frac{1}{n}\sum_{i} V_{i}/I_c$, where $n$ is the number of wavelength positions over which the summation is carried out, and $i$ goes from $-100$ to $-200$~m\AA\/ for the maps in the blue wing, from $100$ to $200$~m\AA\/ in the red wing, and from $600$ to $800$~m\AA\/ for the maps in the far red wing. $I_{c}$ represents the continuum intensity averaged over the SST FOV.

\subsection{IRIS observations} 

The SST data are complemented with IRIS observations taken from 07:58:54 until 11:05:32~UT. The two data sets overlap for 2.5 hours. IRIS recorded medium sparse 2-step rasters, taking spectra in the near ultraviolet band from 2783 to 2834~\AA\/. These spectroscopic measurements also include two spectral regions in the far ultraviolet domain, from 1332 to 1358~\AA\ and from 1389 to 1407~\AA\/. Such a setup allows us to sample the solar atmosphere from the photosphere up to the transition region. The cadence of the spectral observations was 9.3~s per raster step (the total raster cadence is 18.6~s), covering a QS region of $2\times60$~ arcsec$^{2}$.

Slit-jaw images were taken using the \ion{Si}{4} 1400~\AA\ (SJI 1400), \ion{Mg}{2} k 2796~\AA\ (SJI 2796), and \ion{Mg}{2} h wing at 2832~\AA\ (SJI 2832) filters, while keeping solar rotation compensation switched off. In all cases the pixel size is $0\farcs16$. The cadence of the slit-jaw images at 2796~\AA\ is 19~s. Since we were taking one slit-jaw image at 2832~\AA\ instead of every $6^{\text{th}}$ SJI 1400 frame, the final cadences for these two filtergram sequences are respectively 112~s and 19~s (except when an SJI 2832 image is taken, which gives an effective cadence of $~23$~s for SJI 1400). 

The IRIS observations were corrected for dark current, flat-field, geometric distortion, and scattered light. The calibration pipeline also includes wavelength calibrations and subtraction of the background light leak in FUV data. Finally, for the purpose of inverting the radiative transfer equation we performed radiometric calibration to convert the Data Number units of the measured intensities into nW~m$^{-2}$~sr$^{-1}$~Hz$^{-1}$.

\subsection{Inversions of SST data}
\label{inv_sst}

The SST spectropolarimetric data are used to derive the magnetic properties of the solar atmosphere. To this aim, we carried out inversions of the \ion{Fe}{1} 6173 \AA\/ and \ion{Mg}{1}~b$_2$ 5173 \AA\/ lines applying the SIR code \citep{SIR1992}. SIR solves the radiative transfer equation under the assumption of local thermodynamic equilibrium (LTE) and returns the temperature, velocity, magnetic field strength, inclination, and azimuth angles along the line of sight (LOS). LTE inversions of the \ion{Mg}{1}~b$_2$ line can be considered reliable as long as we are interested only in the magnetic field strength in the upper photosphere \citep{delaCruzRodriguezetal2012}.

\begin{figure*}[!t]
	\centering
	\includegraphics[width=0.95\textwidth]{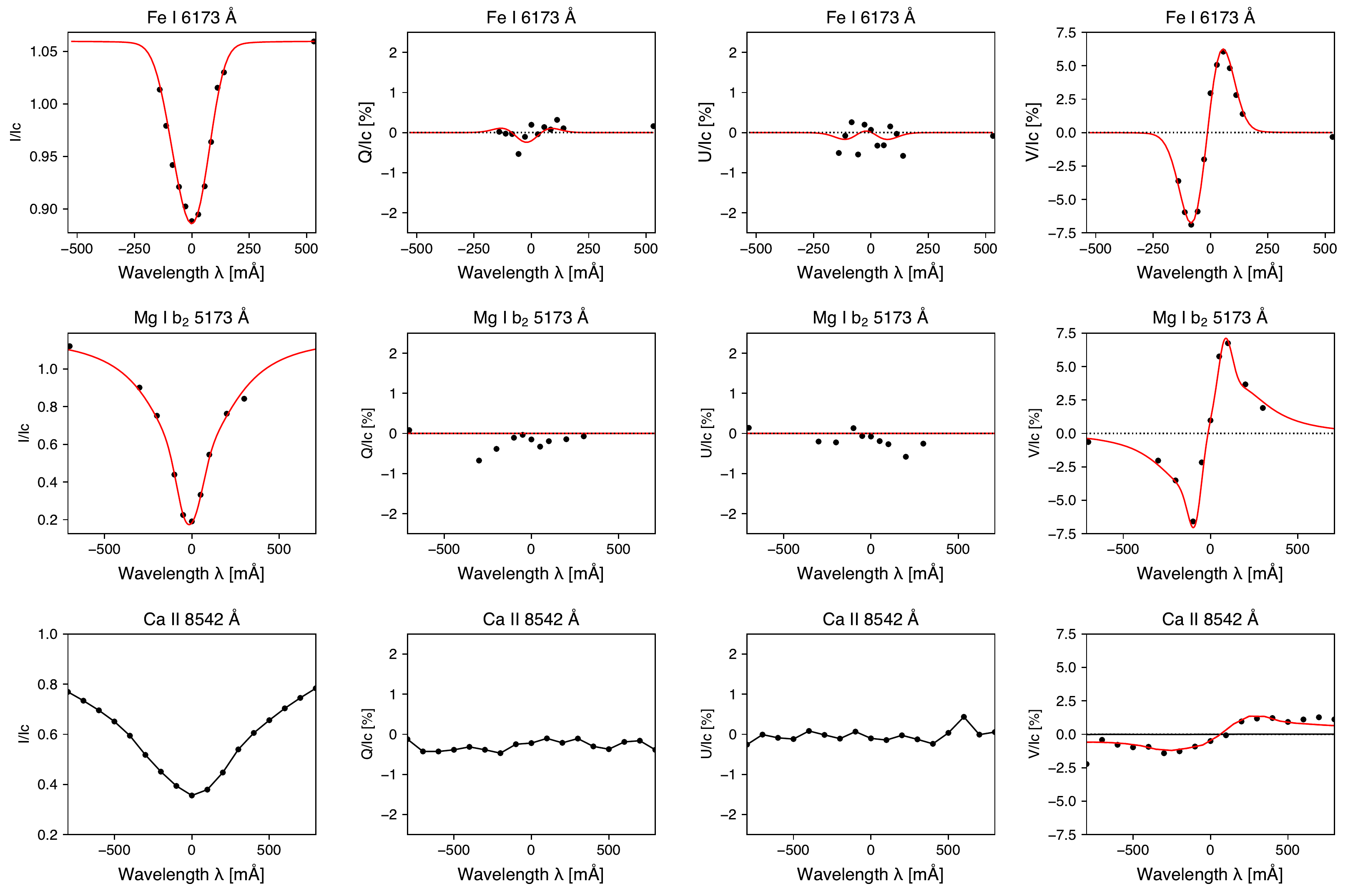}
	%	\vspace*{2.5em}
	\caption{An example of the typical Stokes profiles detected within one of the footpoints of the emerging magnetic bipoles. The spectral lines were observed at the location marked with the orange dot in the magnetograms of Figure \ref{hmi}. From left to right, the black dotted and solid curves show the observed Stokes $I$, $Q$, $U$ and $V$ profiles in the \ion{Fe}{1} 6173 \AA\/ line (upper row), \ion{Mg}{1} b$_{2}$ 5173 \AA\/ (middle row), and \ion{Ca}{2} 8542 \AA\/ (bottom row). The red curves show the best-fit profiles obtained from the SIR inversions and from the WFA.\newline}
	\label{fig_in_inv}
\end{figure*}

For both spectral lines we used a simple one-component model atmosphere and two cycles. In the first cycle, we inverted the \ion{Fe}{1} line considering one node for magnetic field strength, inclination, azimuth and LOS velocity, while having two nodes for temperature. We set the magnetic filling factor to one. Local stray-light contamination is determined for each pixel by averaging the Stokes $I$ profiles in a $1\arcsec$ wide box centered on the pixel, as recommended by \cite{OrozcoSuarezetal2007}. These are reasonable assumptions considering the high spatial resolution of our observations. In the second cycle, the number of nodes for magnetic field strength and LOS velocity were increased to two, in order to account for the asymmetries between the blue and red lobes of Stokes $V$. For the \ion{Mg}{1}~b$_2$ inversions we considered three nodes in temperature. As initial guess model atmosphere we used the Harvard Smithsonian Reference Atmosphere \citep[HSRA;][]{Gingerichetal1971}.

\begin{figure*}[!t]
	\centering
	\includegraphics[width=0.95\textwidth]{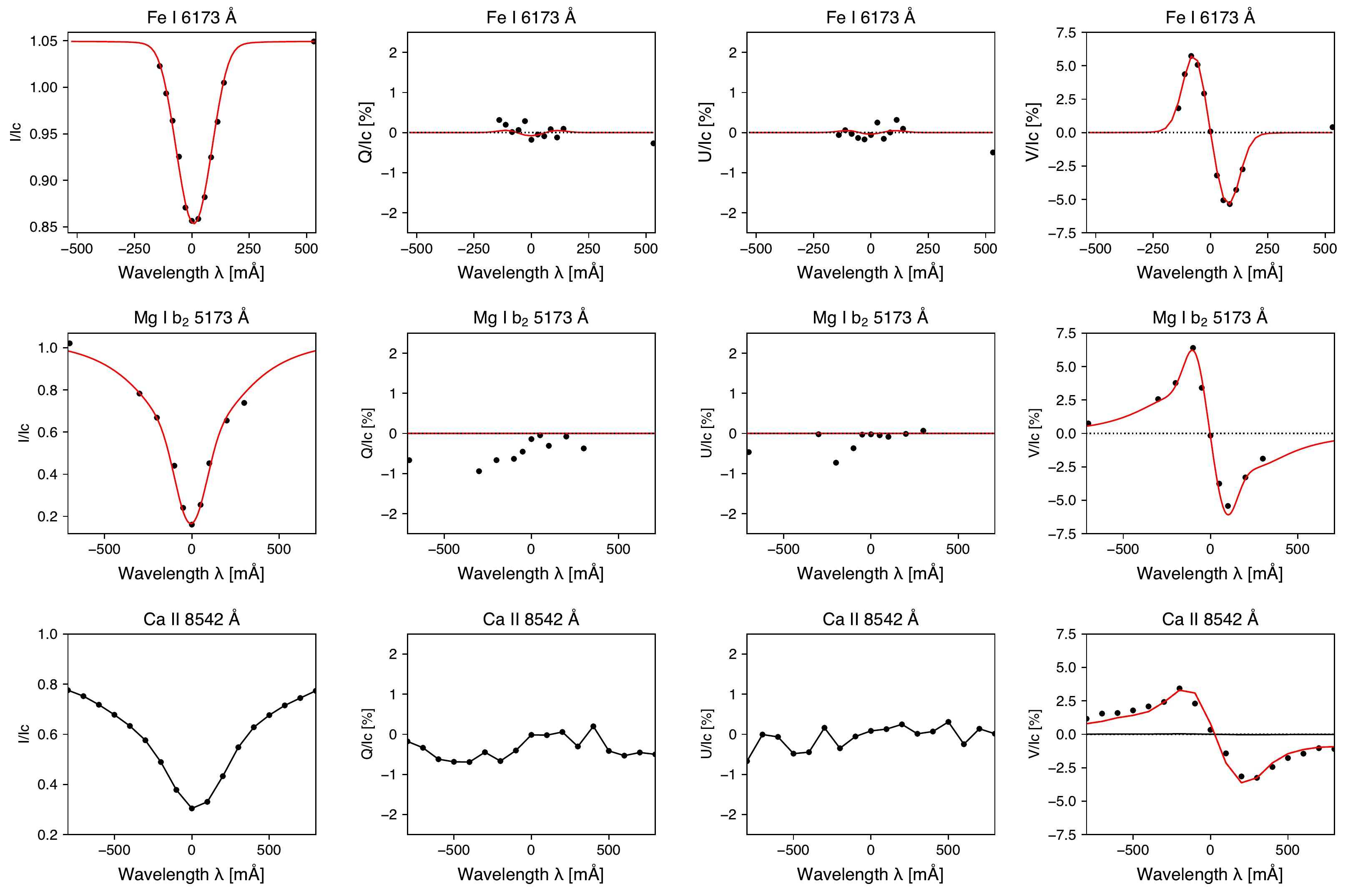}
	%	\vspace*{2.5em}
	\caption{Same as Fig. \ref{fig_in_inv}, but for an NW element. The pixel from which the spectral profiles originate is marked with a violet dot in Figure \ref{hmi}. \newline}
	\label{fig_ne_inv}
\end{figure*}

To avoid expensive non-LTE inversions without clear polarization signals in all four Stokes parameters, we use the weak-field approximation \citep{LandideglInnocenti1992} to infer longitudinal fields from the \ion{Ca}{2} 8542 measurements:

\begin{equation} 
	V(\lambda) = - \phi\ C \frac{\partial I(\lambda)}{\partial \lambda}, 
\end{equation}

\noindent where $V(\lambda)$ is Stokes $V$, $\phi=fB\cos\gamma$ represents the longitudinal flux density, $f$ is the magnetic filling factor, $B$ is the field strength, $\gamma$ is the field inclination with respect to the LOS, $C=4.67 \times 10^{-3} \, g_{\rm eff} \lambda_0^2$, $\lambda_0$ is the central wavelength (in \AA\/), and $g_{\rm eff}$ is the effective Land\'e factor. The longitudinal magnetic flux density in a given pixel can be obtained from a least-squares minimization:

\begin{equation} 
	\frac{\partial}{\partial \phi} \left[ \sum_{i} \left( V_{i} + \phi\ C \frac{\partial I}{\partial \lambda_{i}} \right) \right] = 0, 
\end{equation}

\noindent where the summation extends over all the wavelength positions $i$. According to \citep{MartinezGonzalezBellotRubio2009}, $\phi$ can be expressed from the previous equation as:

\begin{equation} 
	\phi = - \frac{\sum_{i} \frac{\partial I}{\partial \lambda_{i}} V_{i}}{C \sum_{i} (\frac{\partial I}{\partial \lambda_{i}})^{2}}, 
\end{equation}
 
\noindent which can be used to estimate the longitudinal magnetic field assuming the filling factor to be equal to one. This is a reasonable assumption for high-resolution observations. The $i$ summation is carried out within $\pm 200$~m\AA\/ around the core of the line. In this way we suppress contamination from photospheric signal as much as possible. The reason is that the \ion{Ca}{2} line wings are formed in the photosphere, and the \ion{Ca}{2} core in the chromosphere \citep{Pietarila2007, QuinteroNoda2016}. 
Figures \ref{fig_in_inv} and \ref{fig_ne_inv} show examples of observed and synthetic Stokes profiles inside one of the footpoints of an emerging IN bipole and a NW magnetic element, respectively.

\subsection{Inversions of IRIS data}
\label{inv_iris}

One of the main reasons why it remains unclear how IN fields impact the dynamics and energetics of the solar atmosphere is because it is difficult to detect their weak polarization signals to deduce the chromospheric magnetic properties. However, high sensitivity observations and the introduction of machine-learning methods coupled to state-of-the-art inversion codes have provided a breakthrough in our chromospheric diagnostic capabilities. 

To infer the thermodynamical properties of the solar atmosphere inside IN regions, we applied the newly developed IRIS$^{2}$ code \citep{SainzDaldaetal2019} to the IRIS \ion{Mg}{2} h and k and \ion{Mg}{2} triplet lines. IRIS$^{2}$ is an inversion code\footnote{The IRIS$^{2}$ code is publicly available in the IRIS tree of SolarSoft. For more details about the code and the installation see \url{https://iris.lmsal.com/iris2}.} which recovers the thermodynamics as a function of the optical depth in the chromosphere and high photosphere employing the k-means clustering method to build a database of spectral profiles and the corresponding atmospheric models. The k-means algorithm \citep{Steinhaus1957, MacQueen1967} is used to calculate representative profiles (RPs) as the average of observed profiles from different data sets that have a similar shape. 

These RPs were inverted with the STiC code\footnote{STiC is publicly available to the community and can be downloaded from the author’s Web site at \url{https://github.com/jaimedelacruz/stic}.} \citep{delaCruzRodriguezetal2016, delaCruzRodriguezetal2019}. STiC is based on the RH code \citep{Uitenbroek2001} and solves the equations of statistical equilibrium and radiative transfer. It takes into account non-LTE lines and continua including the effects of partial redistribution of scattered photons in strong resonance lines. Inversions are performed assuming plane-parallel geometry and statistical equilibrium ionization to determine the atomic level populations. In that way, a database of $\sim$50,000 observed and inverted IRIS \ion{Mg}{2} h and k spectral profiles was created. This database contains representative model atmospheres (RMAs) derived from the inversions of the observed RPs, the corresponding inverted profiles (inverted RPs), and their response functions (RFs, see \citealt{DelToroIniestaRuizCobo} and references therein).

IRIS$^{2}$ analyzes an IRIS observation and looks for the closest RP in the database. Thus, for each observed profile, the code assigns the model atmosphere resulting from the inversion of the closest RP to that observed profile. The uncertainties in the physical parameters of the model are calculated using the RFs associated with the closest inverted RP in the database. By using this database, IRIS$^{2}$ reduces the time needed to invert a map of the size of an active region by a factor of $10^{5}$-$10^{6}$, while keeping the physical information of the results through the link between the RMAs, the inverted RPs, and the RFs.

\section{Identification and tracking of internetwork bipolar structures}
\label{sect3}

\begin{figure*}[!t]
	\centering
	\resizebox{1\hsize}{!}{\includegraphics[width=1.0\textwidth]{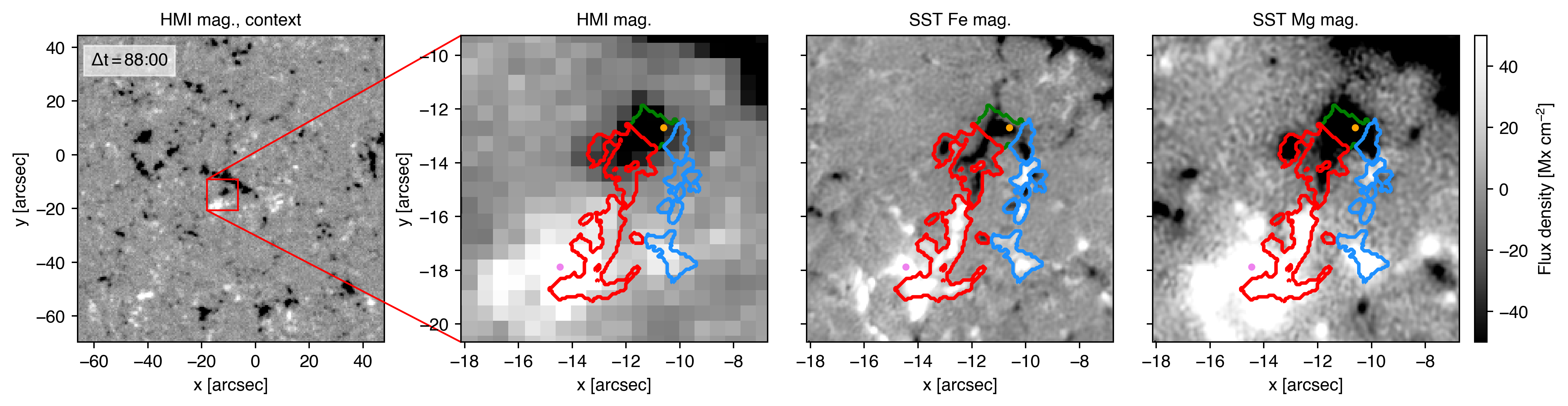}}
	%	\vspace*{2.5em}
	\caption{Still from a movie showing the temporal evolution of newly appeared IN clusters. From left to right: HMI magnetogram showing the QS region where new flux emerges (marked with the red box), a close-up view of the region where only the strongest two footpoints from the three IN clusters are visible, and SST magnetograms taken in the \ion{Fe}{1} and \ion{Mg}{1}~b$_2$ lines revealing the complex magnetic configuration of the clusters. The contours enclose magnetic elements in the photosphere that belong to cluster 1 (green), cluster 2 (red) and cluster 3 (blue). All magnetograms are scaled between -50 and 50~G.\newline
		{\em An animation of this figure is available in the online journal. It runs from $\Delta t$=17:25 to $\Delta t$=147:35, covering $\sim2.2$ hours of observations.}}
	\label{hmi}
\end{figure*}

In order to understand how IN magnetic bipoles/clusters emerge, rise through, and affect the solar atmosphere, we detected and tracked all magnetic elements visible in the FOV, and then focused only on those that appeared in an emerging flux region close to the boundary of a supergranular cell. The selected region (about $5\times7$~arcsec$^{2}$ at the initial stage) produced three bipolar structures in less than one hour. They all emerged on the solar surface as clusters of multiple magnetic loops. Below we describe in detail our method to identify and track magnetic elements appearing within those IN clusters.

\subsection{Criteria for identification}

Stokes $V$ signals are clearly visible in all the lines and are used to study the vertical component of magnetic field at the footpoints of magnetic loops and clusters. With our observations, we detect for the first time circular polarization in the chromosphere within the newly emerging IN flux regions. This allows us to follow the magnetic bipoles from the photosphere up to the chromosphere. We also detect clear linear polarization signals in \ion{Fe}{1}~6173. However, the measurements are not sensitive enough in \ion{Mg}{1}~b$_2$ and \ion{Ca}{2} to detect the Stokes $Q$ and $U$ signals of IN elements. Because of this, the horizontal component of the magnetic field vector is studied only in the LP maps of \ion{Fe}{1}~6173.

As previously described, we used the intensity and circular polarization maps to construct magnetograms in which we automatically identified newly appeared magnetic elements using the downhill method. The tracking is then carried out applying the YAFTA code \citep{WelschLongcope} on those magnetograms. In this way, we can follow individual elements for as long as possible, even if they interact with other same-polarity features during their lifetimes. We considered only pixels with signals above a $3\sigma$ threshold, which is $14$~G, $26$~G, and $60$~G for the \ion{Fe}{1}, \ion{Mg}{1}~b$_2$, and \ion{Ca}{2} magnetograms, respectively. As additional constraints, we used a minimum size of 5 pixels and a minimum lifetime of $\sim2$ minutes (2 frames).

To identify where new magnetic loops appear, we use the LP maps. Since small and weak loops may emerge without a clear preceding LP signal, we also use the intensity and Dopplergram maps for additional information. These maps show if the newly emerging flux patches appear above granules, at their edges, or in intergranular lanes, and if they disturb the solar granulation pattern. To classify an element as a member of a cluster of magnetic features, it must appear in situ within a group of mixed-polarity patches in a relatively small region, emerge more or less at the same time, and move outward along the cluster's magnetic axis.

Taking into account interactions between magnetic elements (merging and fragmentation), we follow the footpoints of the detected loops and classify all elements that fragment from the footpoints as cluster members. Flux patches that merge with the footpoints but appear far from the flux emergence centers are excluded from the clusters. Although the tracking is done automatically, we manually verified the clusters to make sure that all in-situ appeared patches are detected properly. If some small magnetic patches are mistakenly assigned to one cluster instead of another, this would have negligible effects on the results and would not change the conclusions drawn here.

\begin{figure*}[!t]
	\centering
	\resizebox{1\hsize}{!}{\includegraphics[width=1.0\textwidth]{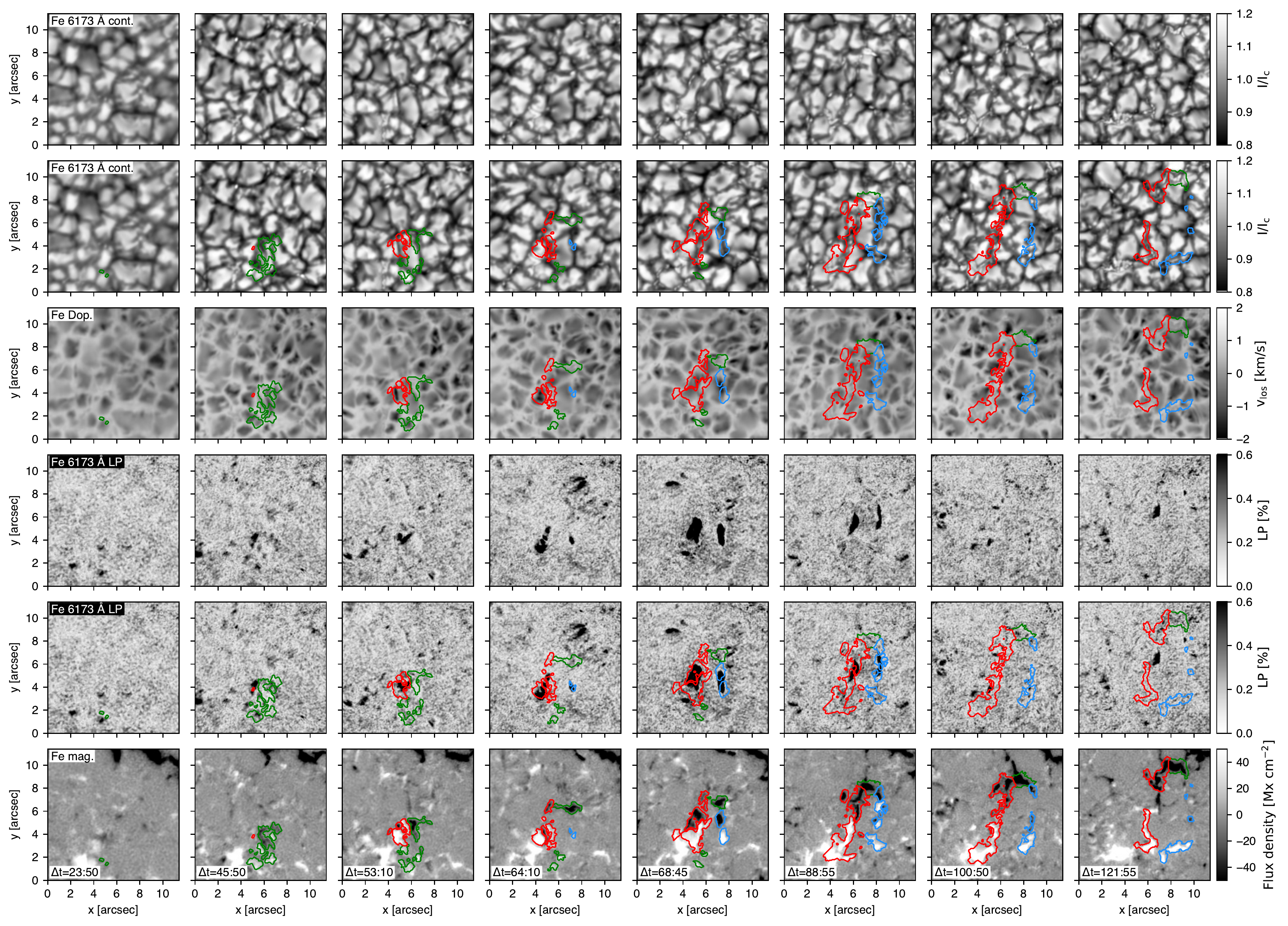}}
	%	\vspace*{2.5em}
	\caption{Emerging IN clusters as seen in the photospheric \ion{Fe}{1} line. Clusters 1, 2 and 3 are enclosed with green, red, and blue contours, respectively. From top to bottom: continuum intensity, continuum intensity with overlaid contours showing the photospheric footpoints, Dopplergrams, linear polarization maps without the contours, linear polarization maps, and longitudinal magnetograms. Note that the linear polarization scale is reversed for the sake of clarity. Starting from $\Delta t$=23:50 [mm:ss], the panels show from left to right the temporal evolution of the footpoints of the clusters that emerge and separate from each other.\newline
		{\em An animation of this figure is available in the online journal, and runs from $\Delta t$=13:45 to $\Delta t$=151:15, covering $2.3$ hours of observations.}}
	\label{fig1}
\end{figure*}

Magnetic features in the \ion{Ca}{2} WFA maps were identified and tracked in a similar fashion as in the Fe and Mg magnetograms, i.e., using YAFTA, a $3\sigma$ threshold, a minimum size of 5 pixels and a minimum lifetime of $\sim2$ minutes. More details about the comparison of longitudinal fields in the \ion{Fe}{1}, \ion{Mg}{1}~b$_2$ and \ion{Ca}{2} lines are given in Sect \ref{blos_fe_mg_ca}.

Information on the upper chromosphere and transition region is provided by the SST H$\alpha$ spectroscopic measurements and the IRIS SJI 2796 and SJI 1400 filtergrams. They show bright points at the footpoint positions and magnetic loops extending between them. Tracking was not necessary for those features, and was therefore not performed.

\section{Results}
\label{sect4}

We observed three IN magnetic clusters emerging at the edges of a supergranular cell. They are located close to each other, and are nested between the surrounding NW fields. The clusters are bound by positive NW patches to the south, and negative ones to the north. They all appeared within 40 minutes. Although having slightly different magnetic axes, the clusters follow the same general orientation, i.e., their strongest negative (positive) polarity footpoints migrate toward the negative (positive) NW fields. 

\subsection{Emergence of IN magnetic clusters in the photosphere}
\label{sect31}

\begin{figure*}[!t]
	\centering
	\includegraphics[width=0.95\textwidth]{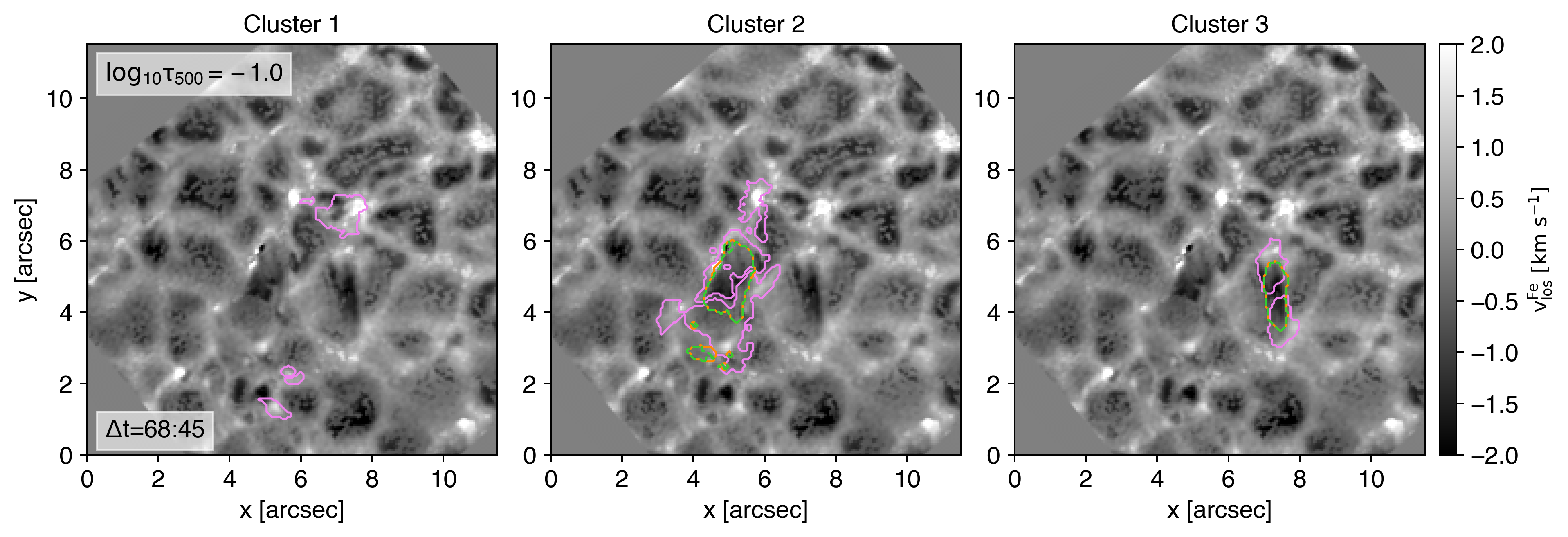}
	%	\vspace*{2.5em}
	\caption{An example of line-of-sight velocity maps derived from the SIR inversions of the \ion{Fe}{1} line. The violet contours show the footpoint positions of the detected IN clusters. The orange and green contours enclose LP patches and upflow pixels within these patches, respectively. The blank corners correspond to pixels for which the inversions were not computed before rotating the velocity maps. \newline
		{\em An animation of this figure is available in the online journal. It runs from $\Delta t$=20:10 to $\Delta t$=139:20, covering $2$ hours of observations.}}
	\label{fig3}
\end{figure*}

\begin{figure}[!t]
	\centering
	\resizebox{1\hsize}{!}{\includegraphics[width=1.0\textwidth]{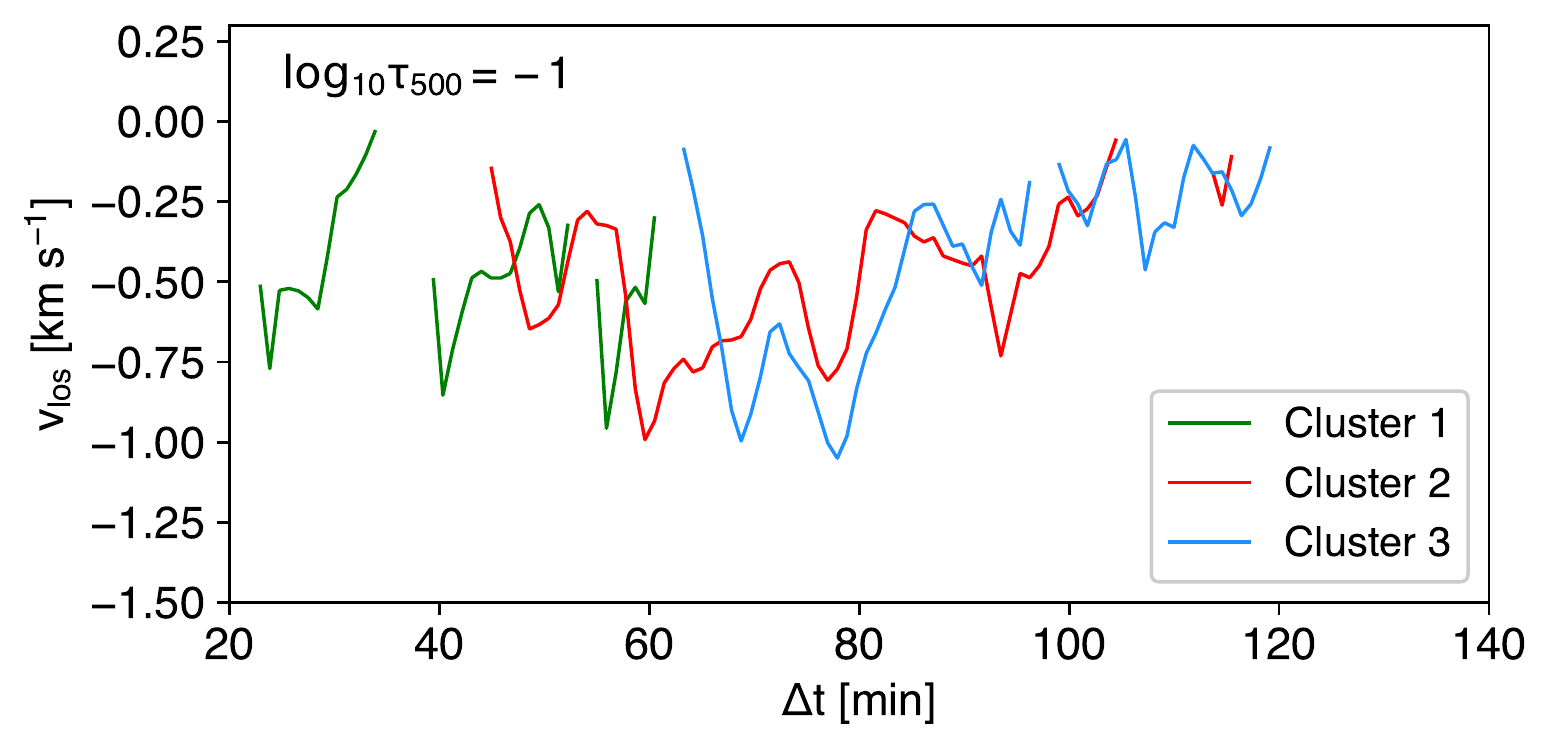}}
	%	\vspace*{2.5em}
	\caption{Upflow velocities within the emerging flux regions. The green (cluster 1), red (cluster 2), and blue (cluster 3) lines give the average upflows in pixels showing clear LP signal. The velocities are derived from the inversion of the \ion{Fe}{1} line, and are only shown for times when upflows are present in patches of significant LP.}
	\label{fig4}
\end{figure}

\begin{figure}[!t]
	\centering
	\resizebox{1\hsize}{!}{\includegraphics[width=1.0\textwidth]{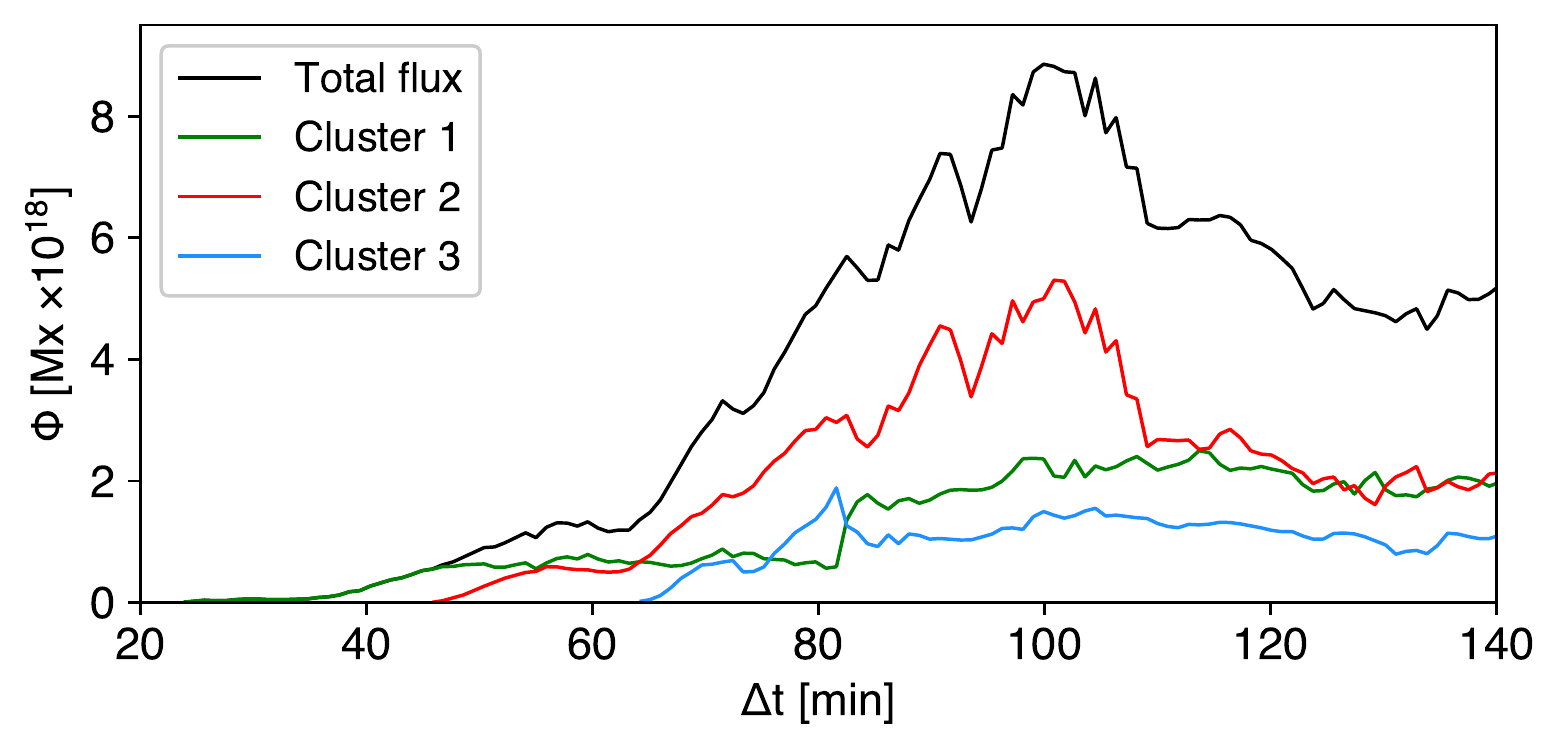}}
	%	\vspace*{2.5em}
	\caption{Temporal evolution of the total unsigned photospheric magnetic flux inside the emerging IN clusters (red, green, and blue curves). The black line shows the sum of the fluxes derived from the \ion{Fe}{1} 6173 circular polarization signals.}
	\label{fig5}
\end{figure}

To give an idea of the spatial extent of the area of flux emergence analyzed in this work, the left panel of Figure \ref{hmi} shows a context magnetogram taken by the Helioseismic and Magnetic Imager \citep[HMI;][]{Scherreretal2012} onboard the Solar Dynamics Observatory \citep[SDO;][]{Pesnelletal2012}. It portrays a QS region with numerous NW elements, and a small IN negative-polarity flux patch inside the red box. This negative patch represents one of the footpoints of the emerging IN clusters. The HMI magnetograms taken every 12 minutes show the Sun with a moderate sensitivity and spatial resolution (1\arcsec). For that reason, they only reveal the NW and the strongest IN features. The second panel from the left shows the same HMI magnetogram, and an enlarged view of the small region indicated by the red square. As can be seen, the negative footpoint at (7\arcsec, 8\arcsec) is clearly resolved, while the positive footpoints are scarcely discernible or appear more as an integral part of the strong positive NW patch located at (4\arcsec, 2\arcsec). From the HMI magnetogram movie however, one may get the impression that the positive magnetic element at (7\arcsec, 3\arcsec) originates from the newly appearing bipolar flux system. By contrast, the SST magnetograms derived from measurements in the \ion{Fe}{1} and \ion{Mg}{1}~b$_2$ lines clearly illustrate a more complex morphology. In particular, they display three separate bipolar clusters of magnetic elements, which are enclosed by the green, red, and blue contours.

\begin{figure*}[!t]
	\centering
	\resizebox{1\hsize}{!}{\includegraphics[width=1.0\textwidth]{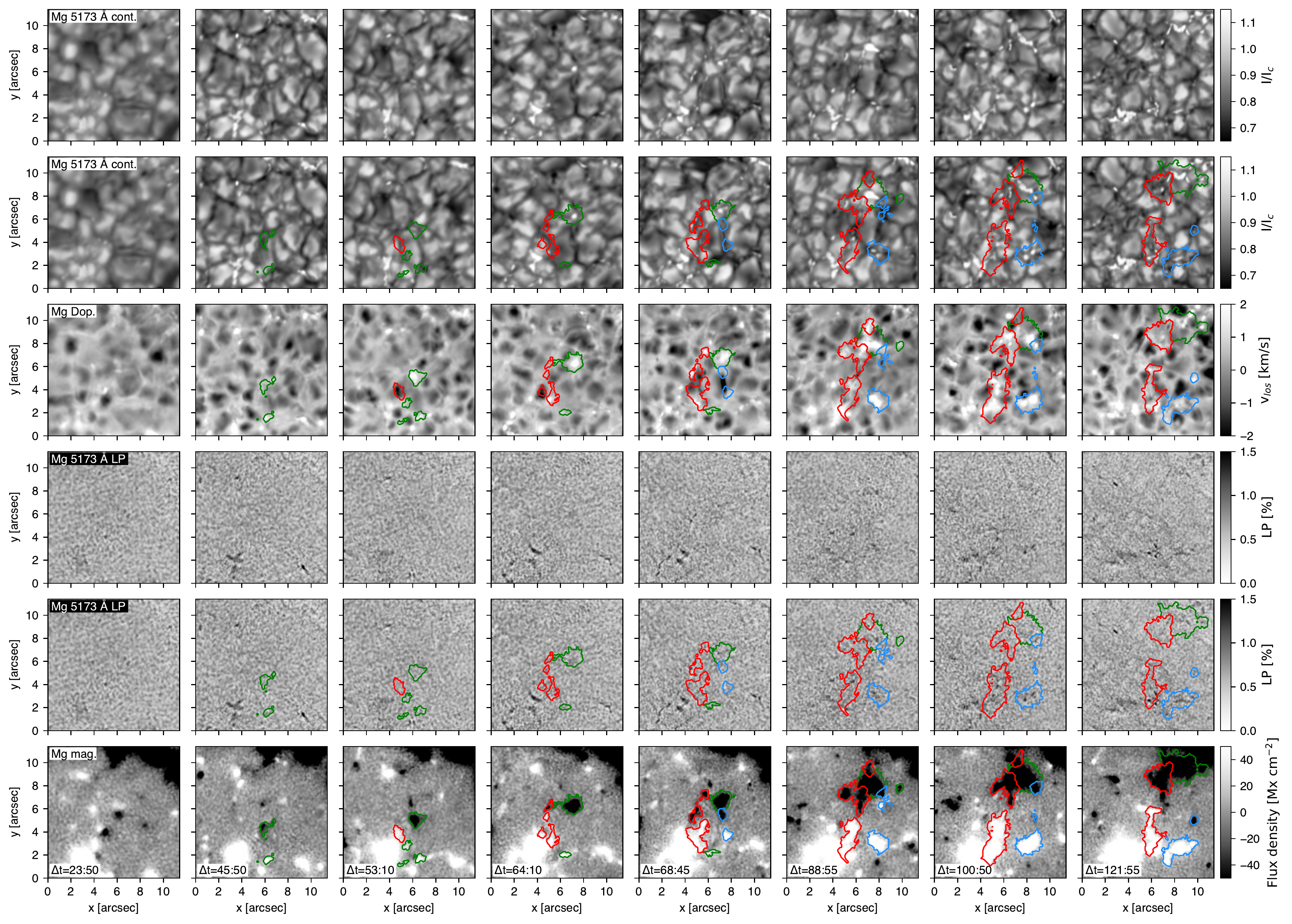}}
	%	\vspace*{2.5em}
	\caption{Same as Fig. \ref{fig1}, but for the \ion{Mg}{1}~b$_2$ line. The contours are based on the \ion{Mg}{1}~b$_2$ magnetograms.\newline
		{\em An animation of this figure is available in the online journal, and runs from $\Delta t$=13:45 to $\Delta t$=151:15, covering $2.3$ hours of observations.}}
	\label{fig2}
\end{figure*}

The first cluster (cluster 1, green contours) appears at $\Delta t$=22:55, and exhibits an episodic pattern of flux emergence. The first hint of IN flux appearing in the photosphere comes from elongated granules, as expected from theoretical models \citep{Cheungetal2007, MartinezSykoraetal2008, TortosaAndreuMorenoInsertis2009} and observed before (e.g. \citealt{Guglielminoetal2010}). This can easily be seen in the intensity and Dopplergram images (Figure \ref{fig1}, first two rows from the top, respectively). At the same time, an LP patch indicating the top of the newly emerging IN loops appears in the Fe LP maps (third row from the top). The LP signal is accompanied by upflows of $-0.5$ to $-0.75$~km~s$^{-1}$. These values are calculated as the mean velocities in pixels having clear LP signals above a $3\sigma$ threshold and negative (upflow) velocities at $\log_{10}\tau_{500}=-1$. Figure \ref{fig3} (left panel) shows the LOS velocity field in and around the emerging flux region. The violet contours enclose the footpoints of the tracked IN clusters, while the orange and green contours mark the boundaries of LP patches and upflow pixels inside, respectively.

In the next frame ($\Delta t$=23:50), the Fe magnetograms show the appearance of new magnetic elements at the flanks of the LP signal. The positive polarity patches immediately start merging with the nearby NW elements, while the negative patches move towards the negative NW structures in the north. The initial emerging phase seems to end at $\Delta t$=30:15, when magnetic elements stop appearing, and those already at the surface continue expanding and separating. The second phase starts at $\Delta t$=34:50, even though the LP signal does not always precede the CP features. Soon after this, the cluster develops into a serpentine-type magnetic structure. The last emerging phase sets off at $\Delta t$=55:00 with upflows of up to $-1$~km~s$^{-1}$. The subsequent photospheric evolution of the flux patches inside cluster 1 is governed by convective flows and interactions with other surrounding magnetic elements. This leads to mixing of positive polarity flux patches with the NW structures in the south. The negative footpoint survives all the interactions and eventually merges with the negative polarity NW fields, as can be seen at the end of the movie accompanying Figure \ref{fig1}.

Cluster 2 appears to the left of cluster 1 at $\Delta t$=44:55, exhibiting the same pattern. First, a clear LP signal appears around (5\arcsec, 4\arcsec), which is followed by mixed-polarity flux patches visible in the magnetograms from $\Delta t$=45:50. New flux patches merge or cancel with the preexisting magnetic structures of cluster 1 and the NW. The second emerging phase is revealed by new LP elements again at (5\arcsec, 4\arcsec), showing up 15 minutes after the first phase. It brings much larger flux features to the surface with upflows of up to $-1$~km~s$^{-1}$. Magnetic elements continue appearing in this region for the next $\sim 60$~minutes, which is not uncommon for IN fields \citep{Gosic2015, Gosicetal2016}. The strongest flux patches in this cluster can be identified and tracked for the entire duration of the magnetogram sequence, despite interacting with other magnetic elements. Since clusters 1 and 2 have the same orientation, many of their flux patches merge together and create more compact and stronger footpoints that eventually become visible in the HMI magnetograms of Figure \ref{hmi}.

The third and smallest bipolar structure in the emerging flux region appears at $\Delta t$=63:15 at (7\arcsec, 4\arcsec). Its magnetic axis is slightly different from that of clusters 1 and 2. However, this small cluster also has its strongest negative-polarity footpoint moving toward the negative NW flux features. The positive-polarity footpoint migrates to the south where it creates a persistent magnetic structure. The strongest detected upflow velocities within cluster 3 slightly exceed $-1$~km~s$^{-1}$.

With time, the three clusters form one large bipolar system. The negative flux patches accumulating in the north give rise to a large magnetic element at (8\arcsec, 8\arcsec), close to the negative polarity NW fields. At the other end of the bipole, positive flux features create an extended flux region, which includes the preexisting positive polarity NW flux in the south.

\begin{figure*}[!t]
	\centering
	\resizebox{1\hsize}{!}{\includegraphics[width=1.0\textwidth]{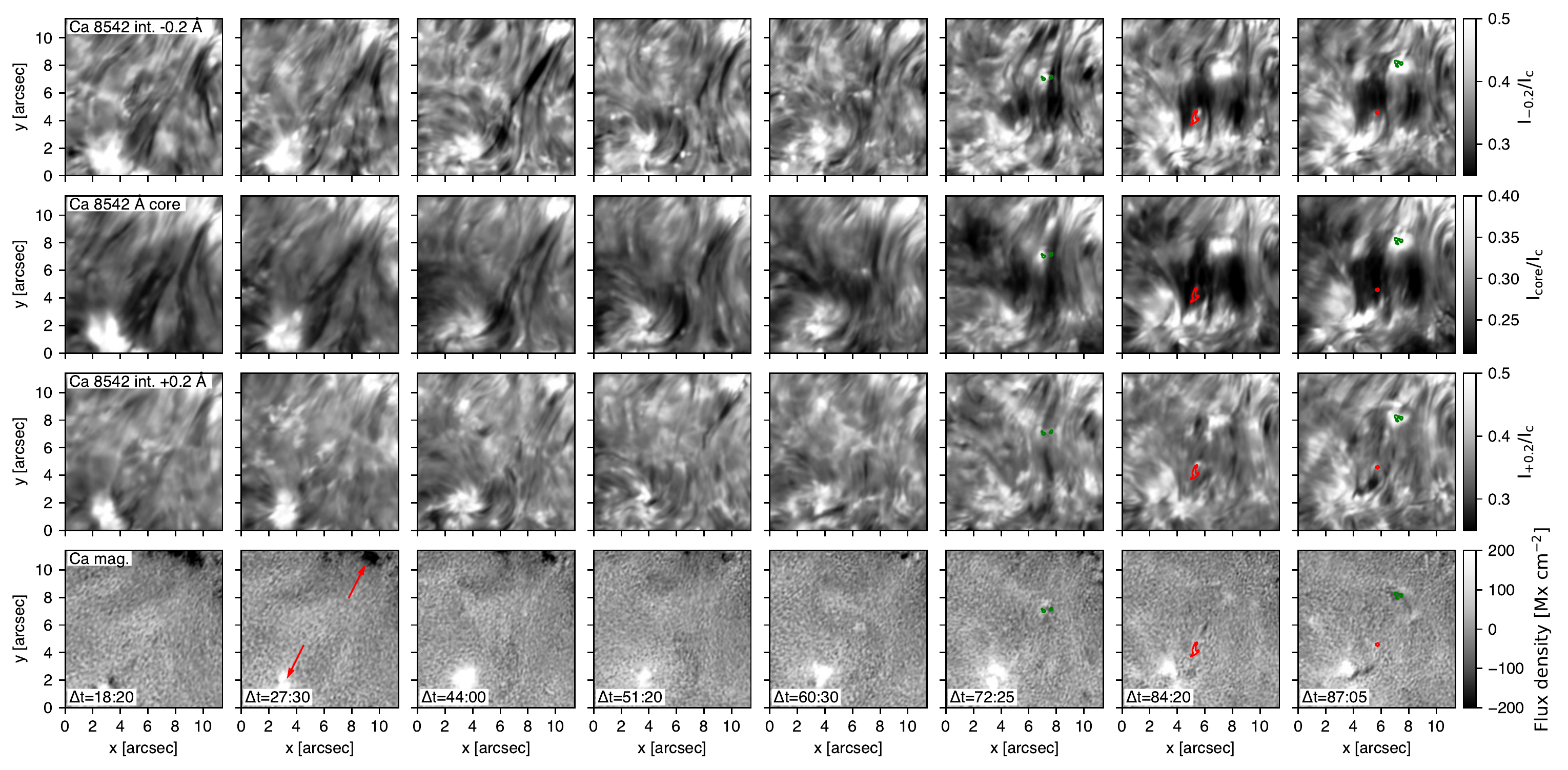}}
	%	\vspace*{2.5em}
	\caption{\ion{Ca}{2} filtergrams showing the temporal evolution of the chromospheric features above the emerging flux region. From top to bottom, the panels display: intensity maps at $-0.2$\AA\/, continuum intensity, intensity at $+0.2$\AA\/, and magnetograms. The contours identify the locations of magnetic elements detected in the \ion{Ca}{2} magnetograms that are co-spatial with the photospheric footpoints of clusters 1, 2, and 3, having the same colors as in Figure \ref{fig1}. Magnetic elements that belong to cluster 2 are not visible in the time steps selected here, but are also prominent in the \ion{Ca}{2} line as can be seen in the movie. Red arrows indicate NW patches. \newline
		{\em An animation of this figure is available in the online journal, and runs from $\Delta t$=13:45 to $\Delta t$=151:15, covering $2.3$ hours of observations.}}
	\label{fig_ca}
\end{figure*}

\begin{figure*}[!t]
	\centering
	{\includegraphics[width=1.0\textwidth]{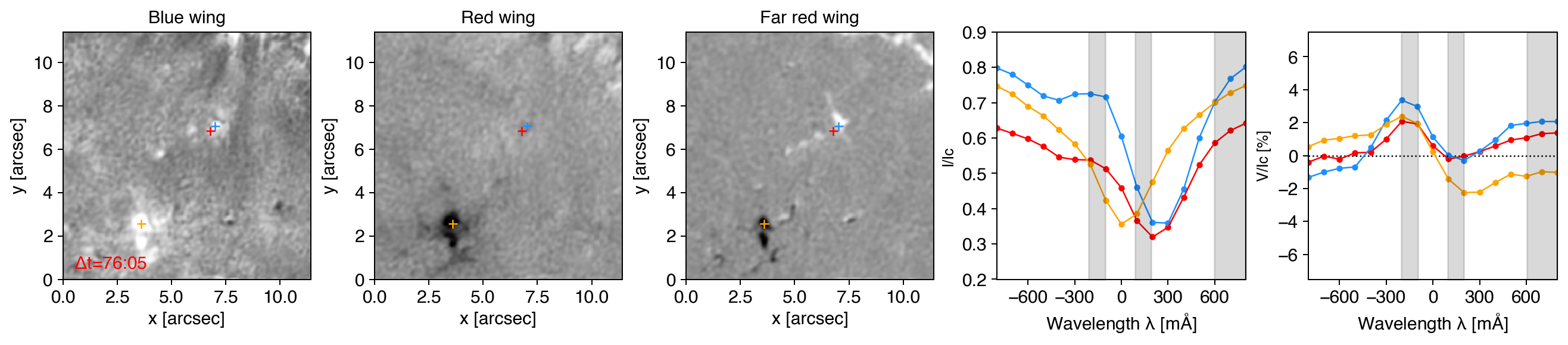}}
	%	\vspace*{2.5em}
	\caption{Intensity and circular polarization profiles at the location of the strongest footpoint at $\Delta t$=76:05. The first three panels from left to right show circular polarization maps in the blue, red and far red wings of the \ion{Ca}{2} line (all scaled between $-1$ and $1\%$). The blue, red and orange crosses mark the pixels for which we show the corresponding spectral profiles in the last two panels. The red and blue Stokes I profiles reveal red-shifted \ion{Ca}{2} lines with emission in their blue wings, which coincides with reversed polarity in the corresponding Stokes V (positive polarity observed at the location of negative polarity photospheric footpoint). The orange curves show as a reference the Stokes profiles in a pixel within the southern NW element. The gray shaded areas indicate the wavelengths used to create the circular polarization maps.
	}
	\label{reversal}
\end{figure*}

Figure \ref{fig4} offers an alternative view of the temporal evolution of these clusters, showing how the average LOS velocities within the emerging IN bipoles vary with time. The curves show the exact timings when upflows and horizontal magnetic fields appear in the photosphere. It can be seen that all three clusters have multiple peaks, suggesting episodic emergence of magnetic flux with maximum upflow velocities of about $-1$~km~s$^{-1}$ at $\log_{10}\tau_{500}=-1$, similar to the values reported by \cite{Guglielminoetal2008}, \cite{Guglielminoetal2012}, and \cite{MartinezGonzalezBellotRubio2009}.

The determination of the total flux of the individual magnetic patches allow us to examine the temporal variations of the magnetic flux inside the clusters. The total flux is calculated separately for each cluster, adding the unsigned fluxes of all the detected individual patches as determined from the measurements obtained in the \ion{Fe}{1} 6173 \AA\/ line (only the CP signals are used). These measurements should give us the most reliable estimates of the total fluxes thanks to their high signal-to-noise ratio. The flux curves are shown in Figure \ref{fig5}. Clusters 1, 2, and 3 bring $2.5\times10^{18}$~Mx, $5.3\times10^{18}$~Mx, and $1.9\times10^{18}$~Mx to the solar surface, respectively. It is important to remember that all the curves are affected by interactions of magnetic elements. This is most obvious between $\Delta t=102$ and $\Delta t=109$~min, when the total flux in cluster 2 quickly drops from $5.3\times10^{18}$~Mx to $2.7\times10^{18}$~Mx. During this time interval, some of the positive polarity flux patches mixed with the NW patch to such an extent that it became impossible to track them further after this time. 

Considering their flux content, the IN clusters studied here do not classify as ephemeral regions (ERs), i.e., bipolar magnetic systems with fluxes in the range $5-30 \times 10^{18}$~Mx and lifetimes of 3--4 hours \citep{Title2000, Chae, Hagenaar2001, Hagenaar2003, Hagenaar2008}. We recall that the definition of ERs is based on the sensitivity and spatial resolution of the observations that were available about two decades ago. Individually, the clusters analyzed here would not satisfy the aforementioned criteria. Moreover, the majority of elements that constitute the clusters---those with fluxes below $10^{18}$~Mx---, would be invisible to instruments with modest spatial resolution such as the Michelson Doppler Imager aboard the SOHO satellite \citep{Scherrer} and HMI. However, taking into account that the total flux inside the emerging region (all three clusters combined) is $8.8\times10^{18}$~Mx at maximum, and that the morphology of the newly emerged flux features resembles that of ERs, we conclude that most likely there is no difference between ERs and clusters of IN elements, and that they all are part of a continuous flux and size distribution of magnetic elements in the QS \citep{Gosic2015}. This idea is consistent with the conclusions of \cite{Parnelletal2009}.

\subsection{Emergence of IN magnetic clusters in the upper photosphere}
\label{sect311}

The loop tops manifested by the LP signals disappear quickly from the photosphere. However, the CP patches, showing the footpoints of magnetic loops, continue to be visible and separate from each other with time. This suggests that the newly emerging loops keep ascending through the solar atmosphere. We do not have enough sensitivity to detect the horizontal fields in the \ion{Mg}{1}~b$_2$ line, but, after the appearance of LP in the \ion{Fe}{1} filtergrams and upflows in the \ion{Mg}{1}~b$_2$ Dopplergrams, it takes only three minutes for the footpoints to appear in the \ion{Mg}{1}~b$_2$ magnetograms at the same positions as in the lower photosphere (similar to the findings of \citealt{MartinezGonzalezBellotRubio2009}). The footpoints are detected separately in the \ion{Fe}{1} and \ion{Mg}{1}~b$_2$ magnetograms, but follow the same evolutionary pattern, as can be seen in Figure \ref{fig2}. However, they do not completely overlap. There are several reasons for that. First, the patches are larger in the \ion{Mg}{1}~b$_2$ line because this line is formed higher than the \ion{Fe}{1} 6173 line \citep{Lites1988} and the magnetic field expands and changes inclination with height. Second, due to different sensitivities and time necessary to scan through the lines, the signals in the \ion{Fe}{1} and \ion{Mg}{1}~b$_2$ lines may show slightly different shapes for the magnetic features. Finally, the different noise levels and the different spatial resolution of the observations may also lead to flux-weighted centers of magnetic elements that do not perfectly overlap. All these can influence the tracking algorithm to generate slightly different contours in the \ion{Fe}{1} and \ion{Mg}{1}~b$_2$ magnetograms.

It is important to note that the footpoints in the upper photosphere often coexist with the horizontal fields in the lower photosphere, for example at $\Delta t$=68:45~min. This suggests that although some of the loops have already reached the upper photosphere, the clusters continue emerging in the layers below. This favors a continuous structure of the emerging fields over a discrete flux bundle, as suggested by \cite{MartinezGonzalezetal2010}.

\subsection{Magnetic field emergence in the chromosphere}
\label{sect32}

Figure \ref{fig_ca} displays intensity maps in the blue wing, core, and red wing of the \ion{Ca}{2} 8542 \AA\/ line (top three rows). The bottom row reveals the temporal evolution of longitudinal magnetic fields in the chromosphere. Before the emergence of the IN clusters, there are two locations with strong NW fields (indicated with red arrows). They are visible at all wavelengths across the \ion{Ca}{2} Stokes \textit{I} profiles as extended bright chromospheric structures, that spatially coincide with bright points in the photosphere (see Figures \ref{fig1} and \ref{fig2}). The two NW regions establish the pre-existing magnetic environment in the form of a canopy of field lines connecting the opposite polarity NW patches. The canopy is indirectly outlined by dark fibrils visible in the \ion{Ca}{2} 8542 line center that are expected to follow the magnetic field lines, although misalignments may occur in weakly magnetized regions \citep{delaCruzRodriguezSocasNavarro2011, MartinezSykoraetal2016, AsensioRamosetal2017}. 

\begin{figure*}[!t]
	\centering
	\includegraphics[width=1.0\textwidth]{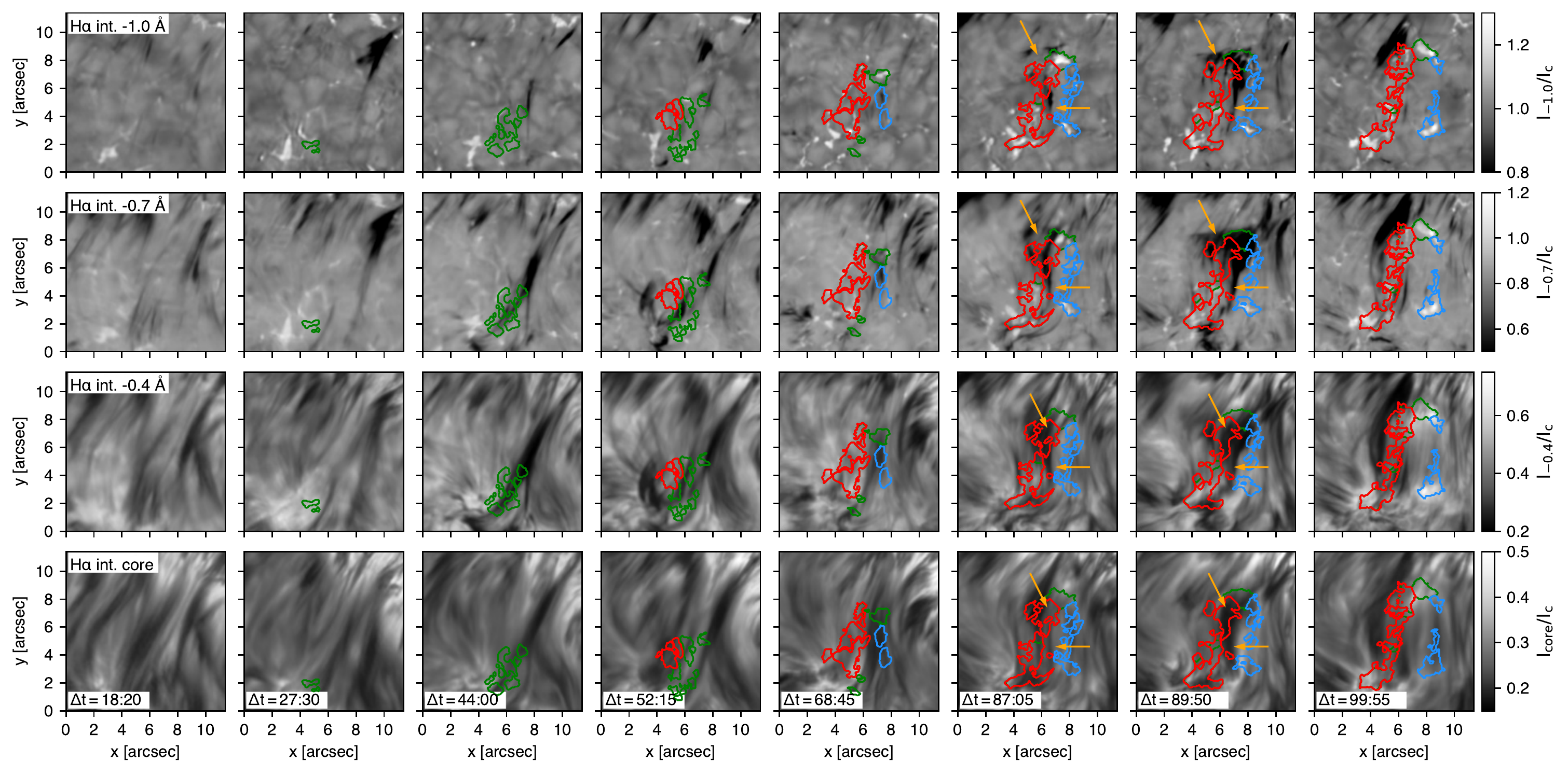}
	%	\vspace*{2.5em}
	\caption{Flux emerging region as seen in the chromospheric H$\alpha$ line. Panels show from top to bottom: intensity maps at $-1$, $-0.7$, and $-0.4$~\AA\/, and the core of the H$\alpha$ line. As in the previous figures, the clusters are enclosed with the green (cluster 1), red (cluster 2), and blue (cluster 3) contours.\newline
		{\em An animation of this figure is available in the online journal, and runs from $\Delta t$=13:45 to $\Delta t$=151:15, covering $2.3$ hours of observations.}}
	\label{fig_ha}
\end{figure*}

The effects of new magnetic loops emerging in the photosphere can be seen very quickly in the chromosphere. We observe that the \ion{Ca}{2} fibrils between the opposite polarity NW patches start bending toward the solar south-west direction $\sim4$ minutes after cluster 1 appears in the photosphere. This deformation becomes evident after $\Delta t$=27:30, both in the blue wing and the core of the \ion{Ca}{2} line, and later also in the red wing, particularly at $\Delta t$=51:20. By this moment, the \ion{Ca}{2} fibrils start delineating loop-like structures between the positive polarity NW flux patch and the newly emerged magnetic elements of clusters 1 and 2. The first hint that we may be seeing the footpoints of cluster 1, ascending through the lower chromosphere, appears around $\Delta t$=60:30 in the blue wing of the line (visible in the movie associated with Figure \ref{fig_ca}). Starting from this moment, we can clearly identify a bright structure at the position of the strong negative polarity footpoint at (7\arcsec, 6\arcsec), that shows a fluctuating intensity for the next $\sim10$~minutes, probably induced by overlying absorbing \ion{Ca}{2} features. At $\Delta t$=72:25, the bright structure becomes easily discernible in the blue \ion{Ca}{2} wing. Two minutes later it shows up in the core of the \ion{Ca}{2} line as well, while pushing the chromospheric \ion{Ca}{2} fibrils to the sides. This marks the moment when a magnetic patch appears in the \ion{Ca}{2} magnetograms around (7\arcsec, 7\arcsec). The polarity is reversed with respect to the photosphere due to emission in the \ion{Ca}{2} blue wing, which is probably a sign of heating in the chromosphere. This positive polarity magnetic element is accompanied by intermittent absorbing fibrils in H$\alpha$ (we will describe the evolution of H$\alpha$ features later). The element disappears completely at $\Delta t$=84:20, followed by extended dark fibrils in H$\alpha$, and is immediately replaced by a negative polarity patch at the same position as the photospheric footpoint (completely visible by $\Delta t$=87:05). 

\begin{figure*}[!t]
	\centering
	\includegraphics[width=0.9\textwidth]{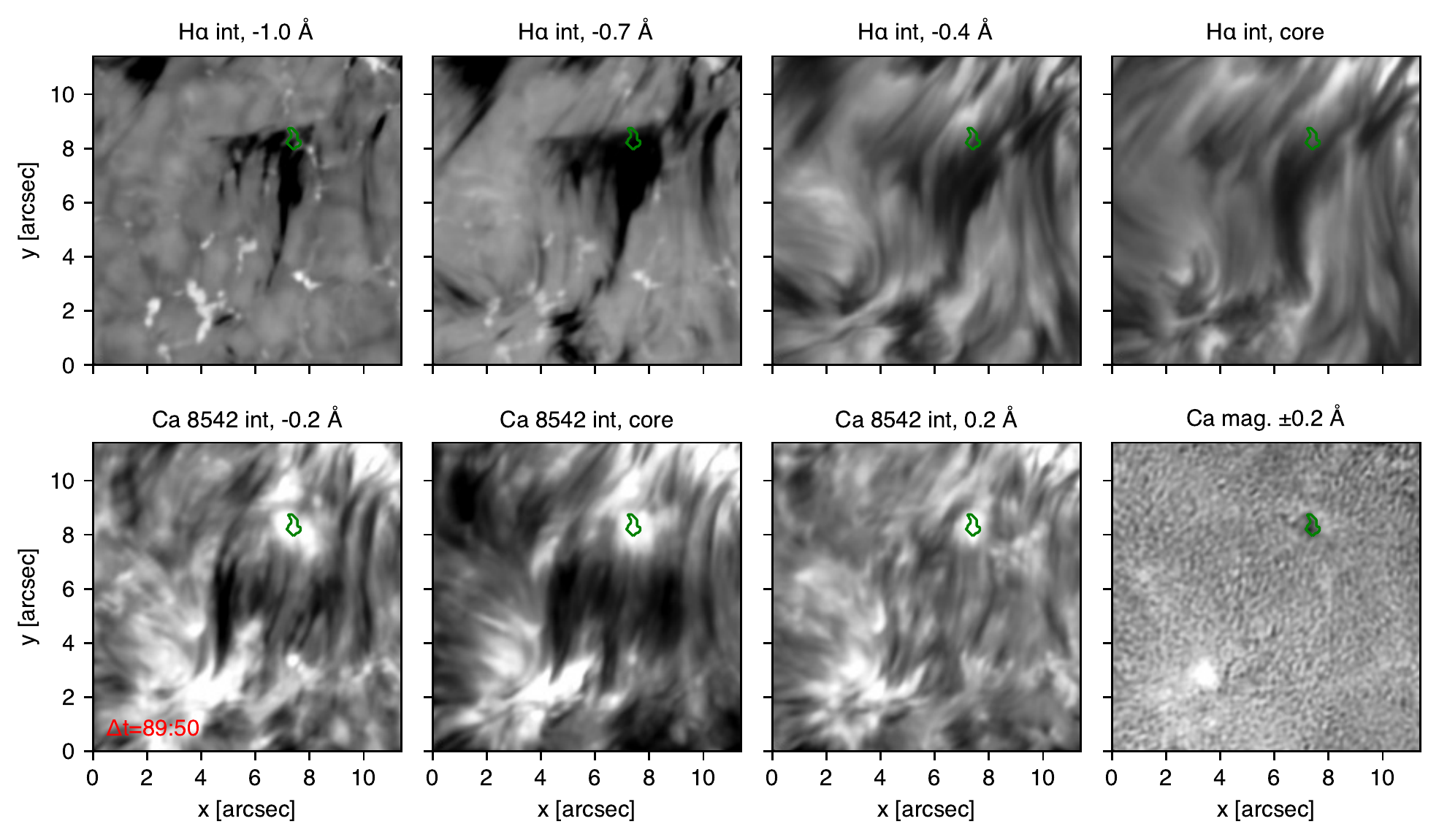}
	%	\vspace*{2.5em}
	\caption{Chromospheric fibrils above the emerging IN flux region at $\Delta t$=89:50. \textit{Upper row}: Blue-shifted absorption H$\alpha$ features are rooted in the footpoints of the newly emerging clusters and also in the positive-polarity NW structure in the south. \textit{Bottom row}: \ion{Ca}{2} filtergrams at $-0.2$~\AA\/, the line core, and $0.2$~\AA\/. The last panel is the \ion{Ca}{2} magnetogram revealing the moment when the negative polarity IN footpoints appear in the chromosphere (green contour).\newline}
	\label{ca_ha_int}
\end{figure*}

We interpret this sequence of events as follows; when the IN loop footpoints reach the chromosphere and start interacting with the pre-existing canopy fields, the interplay between these fields leads to heating, which is observed as emission in the \ion{Ca}{2} blue line wing. The emission causes an apparent polarity reversal of the footpoint (as in \citealt{SainzDalda2007}). An example is presented in Figure \ref{reversal}, which shows the average circular polarization maps in the blue, red and far red wing of the \ion{Ca}{2} line along with the Stokes I and V profiles emerging from three pixels marked with crosses. As can be seen, the polarity reversal happens in the blue wing of the line where the intensity profiles turn into emission. The positive patch in the \ion{Ca}{2} magnetograms disappears later and the footpoint turns into a negative flux patch, as the emission fades. Until the end of the observations, dark \ion{Ca}{2} fibrils are regularly observed to connect the negative polarity footpoint formed by magnetic flux of the three clusters with the positive polarity footpoints (mostly from cluster 3) and the NW element. We do not see linear polarization, probably due to insufficient sensitivity in our \ion{Ca}{2} measurements. 

The H$\alpha$ intensity maps show a very similar morphology to the one observed in the \ion{Ca}{2} filtergrams (Figure \ref{fig_ha}). In addition to a chromospheric rosette rooted in the NW patches, we can see long thick fibrils connecting the positive and negative polarity NW fields before the IN clusters arrive at the solar surface. This can be seen at $\Delta t$=44:00. After that, just like in the \ion{Ca}{2} line, the overlying chromospheric H$\alpha$ fibrils start bending as soon as new flux appears. This first stage of the flux emergence seems to trigger a chromospheric eruption visible as a dark fibril in the blue H$\alpha$ wing at $\Delta t$=52:15 on the left side of cluster 2 (3\arcsec, 4\arcsec). The eruption could be caused by thermal and magnetic pressure building up in the NW region, which is compressed from below by the emerging loops \citep{Yangetal2018, Kontogiannisetal2019}. The other possibility is a surge-like phenomenon expected to be observed in the \ion{Ca}{2} and H$\alpha$ lines when new and pre-existing fields reconnect \citep{Guglielminoetal2010, Guglielminoetal2018, Guglielminoetal2019, NobregaSiverioetal2017}. The most interesting event in the H$\alpha$ maps happens at $\Delta t$=87:05 (see orange arrows in Figure \ref{fig_ha}). Long and wide blue-shifted absorption features appear above the negative polarity footpoints, and extend down to the positive polarity NW patch. The observed absorption structures may outline magnetic field lines \citep{Mooroogenetal2017} rising intermittently through the atmosphere \citep{Ortizetal2016} and reconnecting with the ambient fields above. They can be observed for about ten minutes. As in \cite{NobregaSiverioetal2017}, co-spatially to the blue-shifted absorption fibrils, but two minutes later, we detect in the SJI 1400 filtergrams the formation of new bright loop-like structures connecting the footpoints, and also the negative footpoint and positive NW feature (see the animation associated with Figure \ref{iris_movie}). Simultaneously, it is noticeable that the \ion{Ca}{2} fibrils have completely changed their morphology by now, revealing the fibrils rooted also in the footpoints of the clusters and NW fields in the south. This is illustrated in Figure \ref{ca_ha_int}, which shows Ca II and H$\alpha$ chromospheric fibrils above the IN emerging flux region at $\Delta t$=89:50. These fibrils are different from those low-lying loops connecting the NW structures from the beginning of the observations. After the blue-shift event, the \ion{Ca}{2} magnetograms start showing the positive (3\arcsec, 8\arcsec) and negative polarity (7\arcsec, 8\arcsec) footpoints created by the three clusters, suggesting that even the weakest emerging IN fields managed to reach the chromosphere.

\begin{figure*}[!t]
	\centering
	\includegraphics[width=1.0\textwidth]{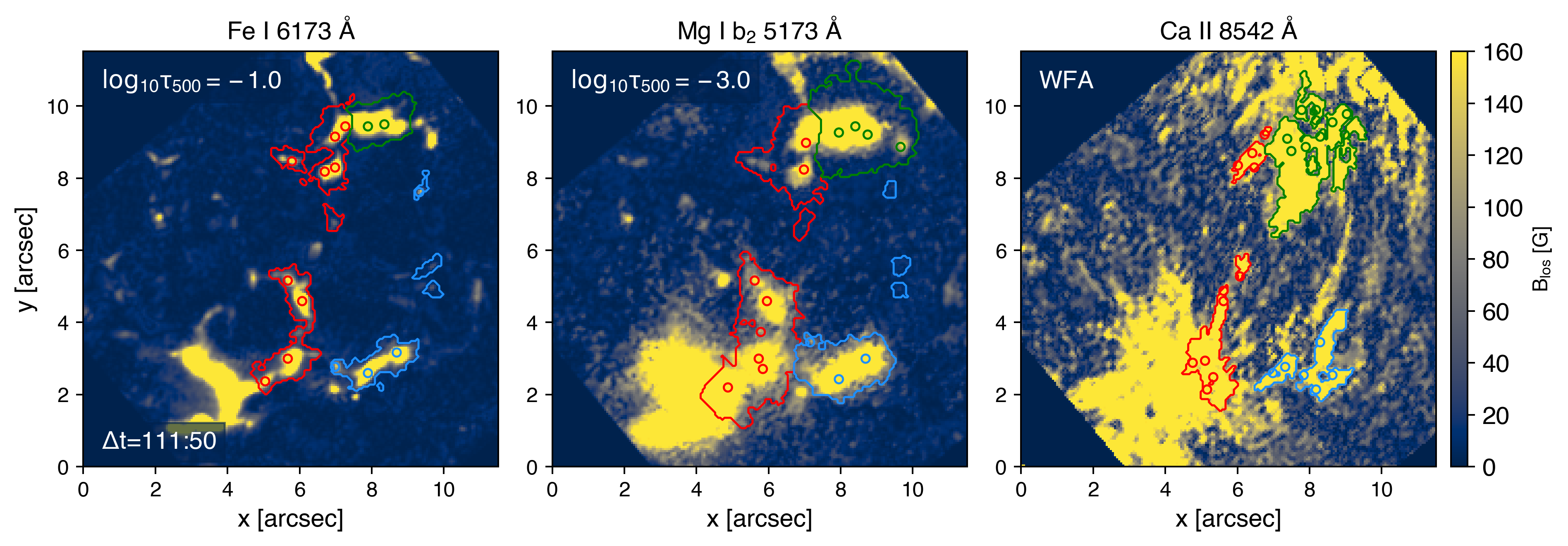}
	%	\vspace*{2.5em}
	\caption{Longitudinal magnetic fields derived from the SIR inversions of the \ion{Fe}{1} and \ion{Mg}{1} b$_2$ lines, and from the WFA applied to the \ion{Ca}{2} measurements at $\Delta t$=111:50. The red, green, and blue contours enclose the footpoints of the newly emerging patches. The contours are defined separately for the three lines. The positions of the local maxima within magnetic elements are marked with circles whose colors match the contours of the corresponding elements. The empty corners correspond to pixels for which the inversions and the WFA were not computed.\newline
		{\em An animation of this figure is available in the online journal. It runs from $\Delta t$=20:10 to $\Delta t$=139:20, covering $2$ hours of observations.}}
	\label{blos_loc_max}
\end{figure*}

\begin{figure}[!t]
	\centering
	\resizebox{1\hsize}{!}{\includegraphics[width=1.0\textwidth]{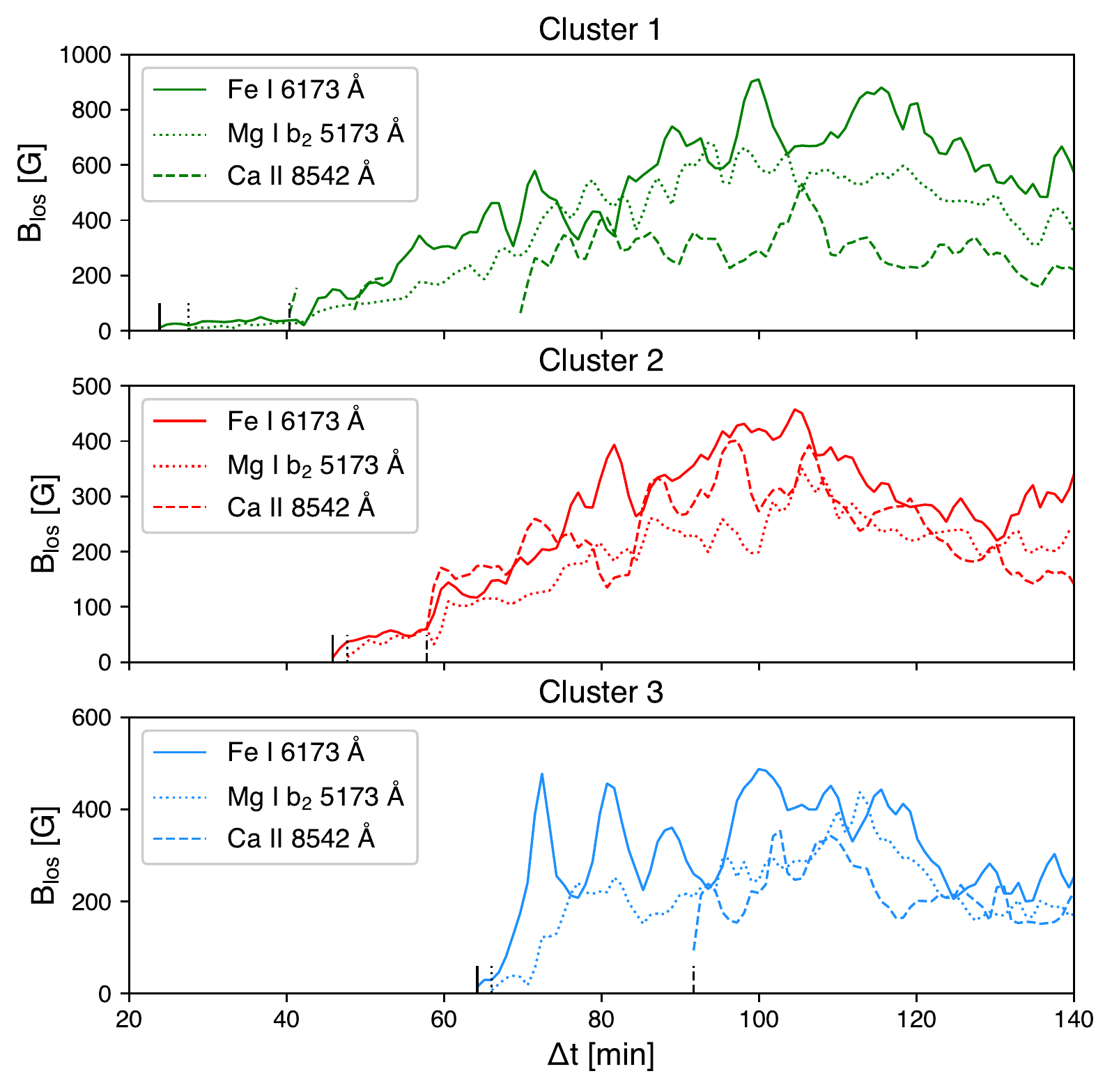}}
	\caption{Temporal evolution of the average longitudinal fields of the newly emerging IN clusters. From top to bottom: Average longitudinal fields for clusters 1, 2, and 3, respectively. The solid curves represent the fields derived from measurements in the \ion{Fe}{1} line, while the dotted and dashed lines represent the fields from the \ion{Mg}{1} b$_2$ and the \ion{Ca}{2} lines, respectively. The vertical solid, dotted and dashed black lines show the moments when the fields appear for the first time in the corresponding spectral lines.\newline}
	\label{blos_core}
\end{figure}

\begin{figure*}[!t]
	\centering
	\resizebox{1\hsize}{!}{\includegraphics[width=1.0\textwidth]{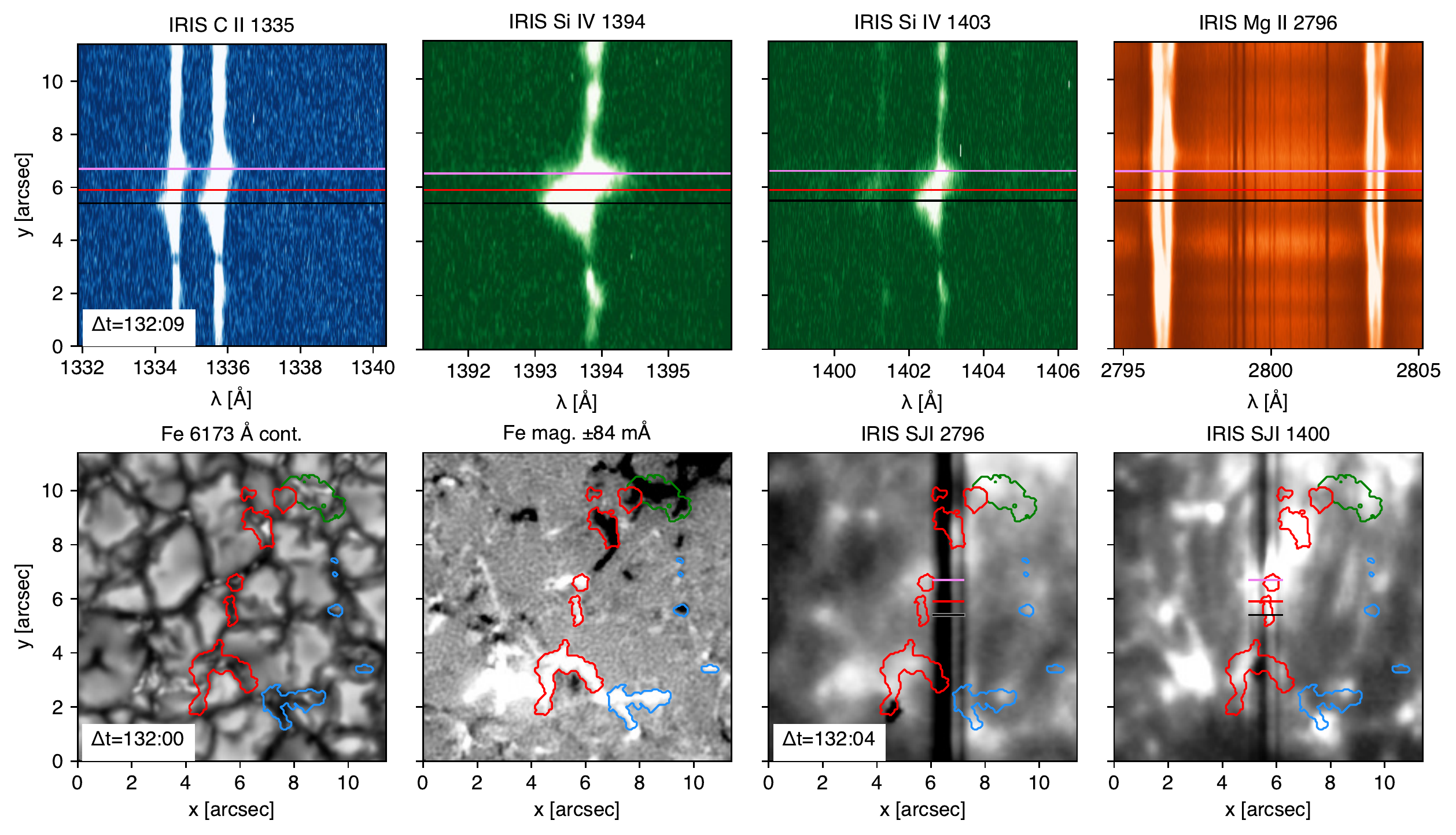}}
	%	\vspace*{2.5em}
	\caption{Temporal evolution of the signals in the IRIS FUV and NUV spectral regions above the emerging IN clusters. \textit{Upper row}: IRIS rasters in the \ion{C}{2} 1335, \ion{Si}{4} 1394, \ion{Si}{4} 1403, and \ion{Mg}{2} 2796 spectral domains. The black, red, and violet horizontal lines mark positions along the slit for which we show spectral profiles later in the manuscript. \textit{Bottom row}: SST continuum intensity maps and magnetograms in the \ion{Fe}{1} line, IRIS SJI 2796, and SJI 1400 intensity maps. The filtergrams display three IN clusters at photospheric and chromospheric heights where many loop-like structures are observed to connect the cluster footpoints and heat the upper solar atmosphere through reconnection with the ambient field lines.\newline
		{\em An animation of this figure is available in the online journal. It runs from $\Delta t$=35:45 to $\Delta t$=151:15, covering $1.9$ hours of observations. The black, red, and violet vertical lines are omitted from the animation for clarity.}}
	\label{iris_movie}
\end{figure*}

\begin{figure*}[!t]
	\centering
	\resizebox{1\hsize}{!}{\includegraphics[width=1.0\textwidth]{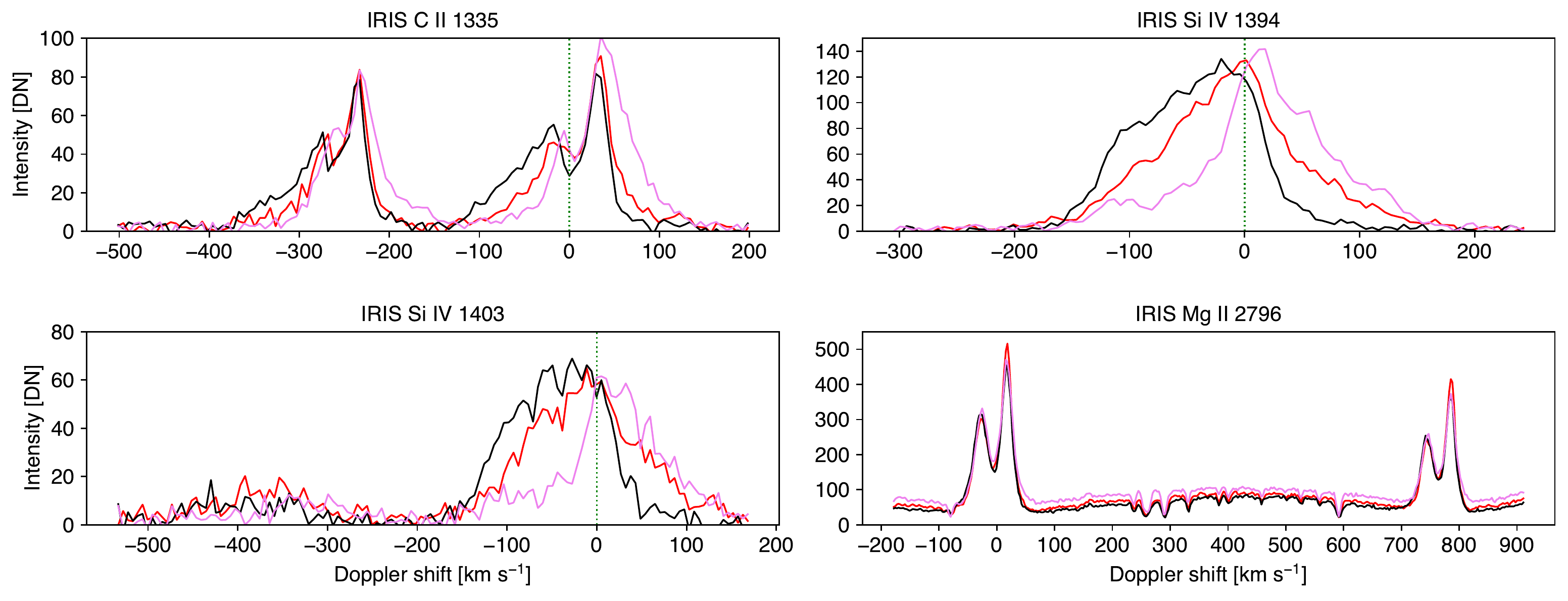}}
	%	\vspace*{2.5em}
	\caption{Selected spectral profiles in the IRIS \ion{C}{2} 1335 (upper left), \ion{Si}{4} 1394 (upper right), \ion{Si}{4} 1403 (lower left), and \ion{Mg}{2} h\&k (lower right) lines at different locations along the slit. The far-UV profiles show strong blue and red shifts during the maximum emission, indicating plasmoid-mediated magnetic reconnection between the upward moving IN magnetic fields (suggested by \ion{Mg}{2} h\&k profiles) and the ambient NW fields.\newline
	}
	\label{iris_profiles}
\end{figure*}

\subsection{Magnetic field within IN clusters}
\label{blos_fe_mg_ca}

To determine the longitudinal field of the newly appeared clusters, we used SIR inversions of the \ion{Fe}{1} and \ion{Mg}{1} b$_2$ lines, and the WFA in the \ion{Ca}{2} line. Since magnetic features may change inclination and shape with height, we do not calculate the field from both lines in the same spatial pixels. Instead, for each magnetic element in each line, we determine the position of the local field maximum, take all the pixels within a radius of 3 pixels and calculate the average LOS magnetic field there. This gives the longitudinal field in the core of the magnetic features. An example of longitudinal field maps in the \ion{Fe}{1}, \ion{Mg}{1} b$_2$, and \ion{Ca}{2} lines with the identified local maxima is shown in Figure \ref{blos_loc_max} and the accompanying movie. To compare the fields in \ion{Fe}{1}, \ion{Mg}{1} b$_2$ and \ion{Ca}{2}, we consider only the pixels that have magnetic signals in all the lines. Therefore, it is reasonable to assume that they belong to the same magnetic structures. We can see here that the \ion{Fe}{1} and \ion{Mg}{1}~b$_2$ field maps reveal elongated and roundish magnetic patches in the photosphere and chromosphere concentrating along the intergranular lanes, respectively. On the other hand, the \ion{Ca}{2} WFA maps show magnetic structures co-located with the tracked footpoints and NW elements, and also loop-like features between the footpoints.

Figure \ref{blos_core} shows the temporal evolution of the longitudinal field inside the newly emerging IN clusters. The vertical solid, dotted and dashed black lines mark the moments when we detect magnetic signals for the first time in the \ion{Fe}{1}, \ion{Mg}{1} b$_2$, and \ion{Ca}{2} maps, respectively. The first appearance of signals in the \ion{Ca}{2} line may be due to noise considering how weak and intermittent they are. Once the \ion{Ca}{2} magnetic structures become stable and show the same polarity as their photospheric counterparts, we can assume that we are seeing the IN clusters reaching the chromosphere. 

\begin{figure}
	\centering
	\resizebox{1\hsize}{!}{\includegraphics[width=1.0\textwidth]{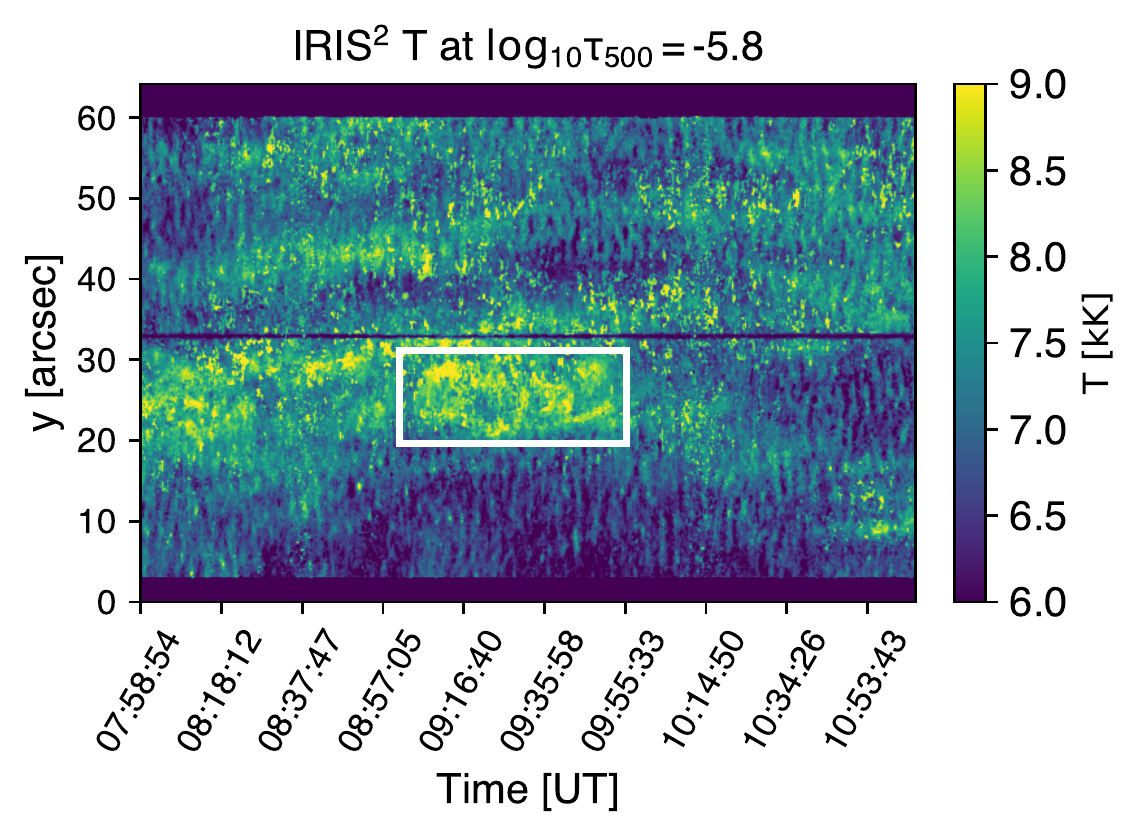}}
	\caption{Temperature maps obtained through IRIS$^{2}$ inversions. The highest temperatures correspond to the IN emerging flux region (indicated by the white box), and NW magnetic elements.}
	\label{mgmapold}
\end{figure}

\begin{figure*}
	\begin{center}
		\resizebox{.49\hsize}{!}{\includegraphics[]{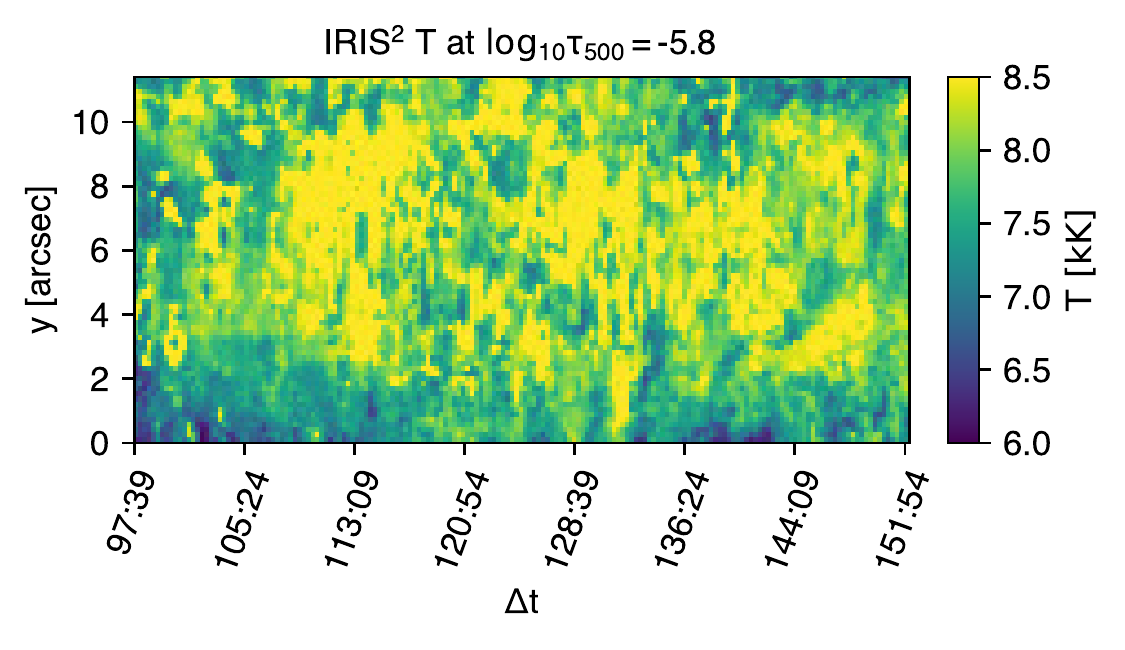}}
		\resizebox{.49\hsize}{!}{\includegraphics[]{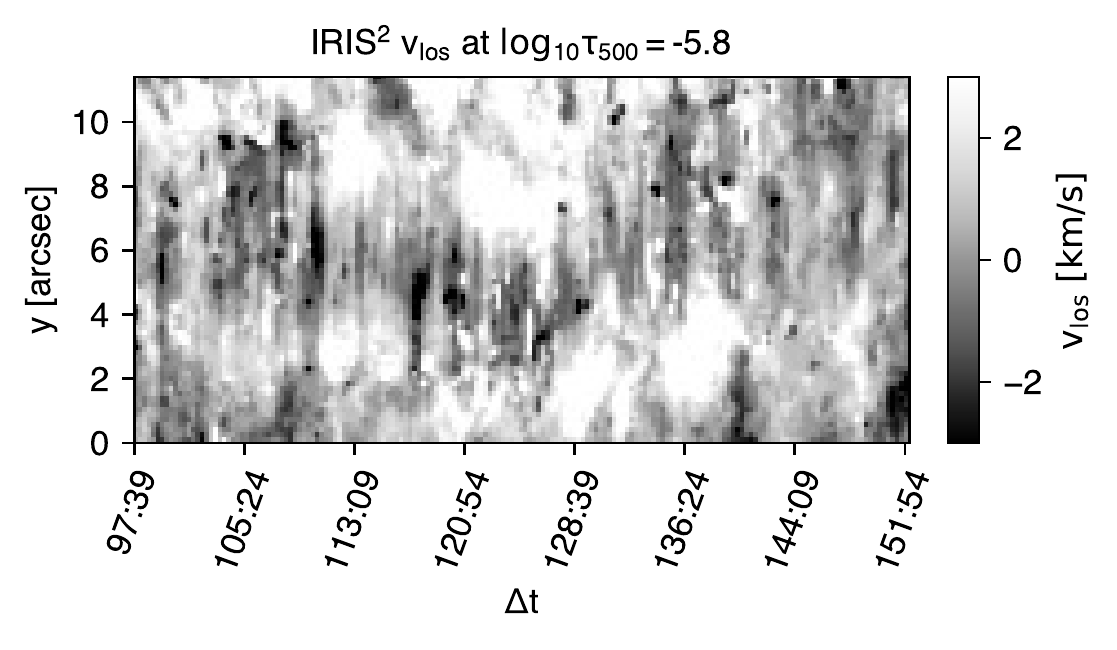}}
	\end{center}
	\vspace*{-1em}
	\caption{IRIS$^{2}$ inversions showing the temperature (left) and LOS velocity (right) maps at $\log_{10}\tau_{500}=-5.8$. The temperature maps shows higher values within the flux emerging region. The LOS velocities reveal upflows around the loop tops and downflows that coincide with the footpoints and NW elements.}
	\label{comparison}
\end{figure*}

The longitudinal fields derived from the SIR inversions of the Stokes profiles in the \ion{Fe}{1} and \ion{Mg}{1} b$_2$ lines (solid and dotted curves, respectively), are very comparable, but the \ion{Mg}{1} b$_2$ line shows weaker fields (100-200~G on average). This is expected since the former line samples the photosphere ($\text{log}_{10}\tau_{500}=-1$), while the latter line is more sensitive to magnetic structures in the upper photosphere (around $\text{log}_{10}\tau_{500}=-3$). The weakest fields are found in the \ion{Ca}{2} line, which samples the chromosphere. The general trend in all clusters is that the longitudinal field increases with time as stronger magnetic features arrive at the solar surface and later accumulate and intensify in the intergranular lanes. The maximum longitudinal fields in clusters 1, 2, and 3 are about 800~G, 450~G, and 500~G at photospheric levels and get weaker with height.

\subsubsection{Uncertainties}

It is clear that our method to select the pixels to compare the longitudinal fields in the SST lines can affect the results. To estimate the uncertainties introduced by the method, we tested how different selection criteria may influence the average longitudinal field values. If we relax the criteria and include all pixels inside the contours defining magnetic patches, i.e., if we take into account also the weakest fields, the average longitudinal fields can decrease by up to 40\% in the \ion{Fe}{1} line and less in the other two lines. The reason is that the photospheric \ion{Fe}{1} line has a better sensitivity and shows weaker fields, thus, we may get a false impression that the magnetic field values are lower in the photosphere than in the upper photosphere or chromosphere. Because of this, we decided to use more strict criteria for the pixel selection, and focus only on the strongest pixels located in the core of magnetic elements for which we can be sure we are studying the same structure. Although the values may change, the shapes of the curves showing the temporal variations of the fields are practically the same for any pixel selection criteria we tested.

\begin{figure}[!t]
	\centering
	\resizebox{1\hsize}{!}{\includegraphics[width=1.0\textwidth]{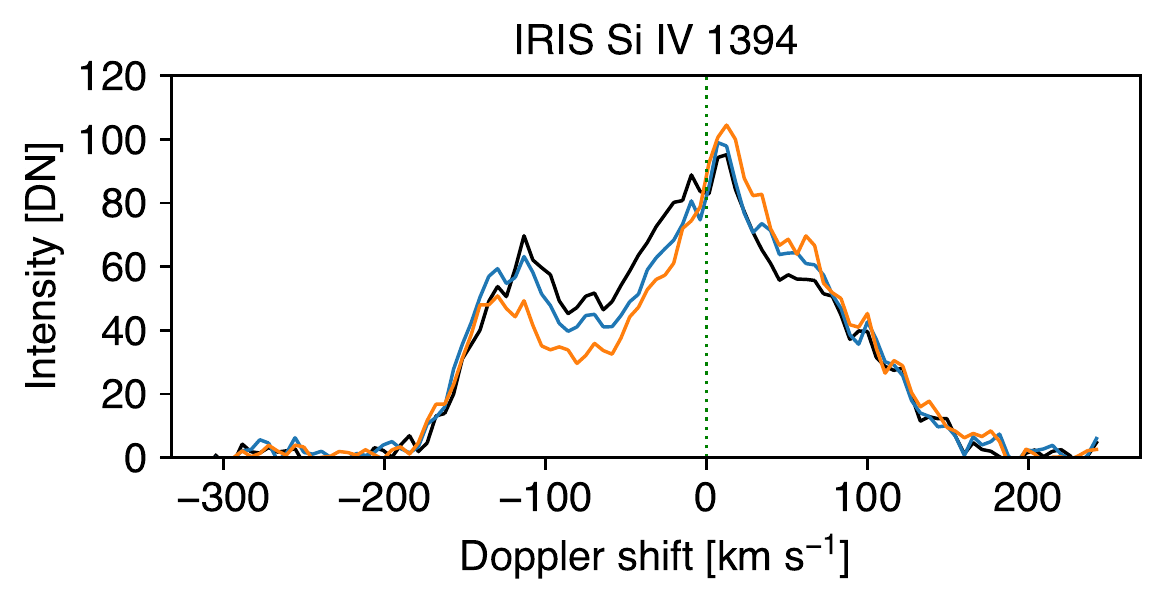}}
	%	\vspace*{2.5em}
	\caption{Examples of double-peaked IRIS \ion{Si}{4} 1394 profiles along the slit, and above the IN emerging flux region. The profiles were recorded at $\Delta t$=132:22.\newline
	}
	\label{iris_profiles2}
\end{figure}

Apart from the noise that can affect the estimates of the magnetic field, the WFA also introduces uncertainties. Using the field values in the \ion{Fe}{1} and \ion{Mg}{1} b$_2$ lines, we determined that the WFA and SIR inversions agree within 10\% for the \ion{Fe}{1} line, and 13\% for the \ion{Mg}{1} b$_2$ line. Having these numbers in mind, we can expect that with much higher noise in the \ion{Ca}{2} observations, the errors can only be larger. In addition, using $\pm200$~m\AA\/ to extract the signal from the core of the \ion{Ca}{2} line has a consequence that we may have a higher noise in the field maps. In principle, we could use $\pm300$~m\AA\/ around the core, but this would increase the photospheric contamination. Therefore, even though we detect a clear signal in the \ion{Ca}{2} line, the estimated fields may not be accurate, while the longitudinal field derived from the SIR inversions of the \ion{Fe}{1} Stokes parameters should be taken as the most reliable.

\begin{figure*}
	\centering
	\resizebox{1\hsize}{!}{\includegraphics[width=1.0\textwidth]{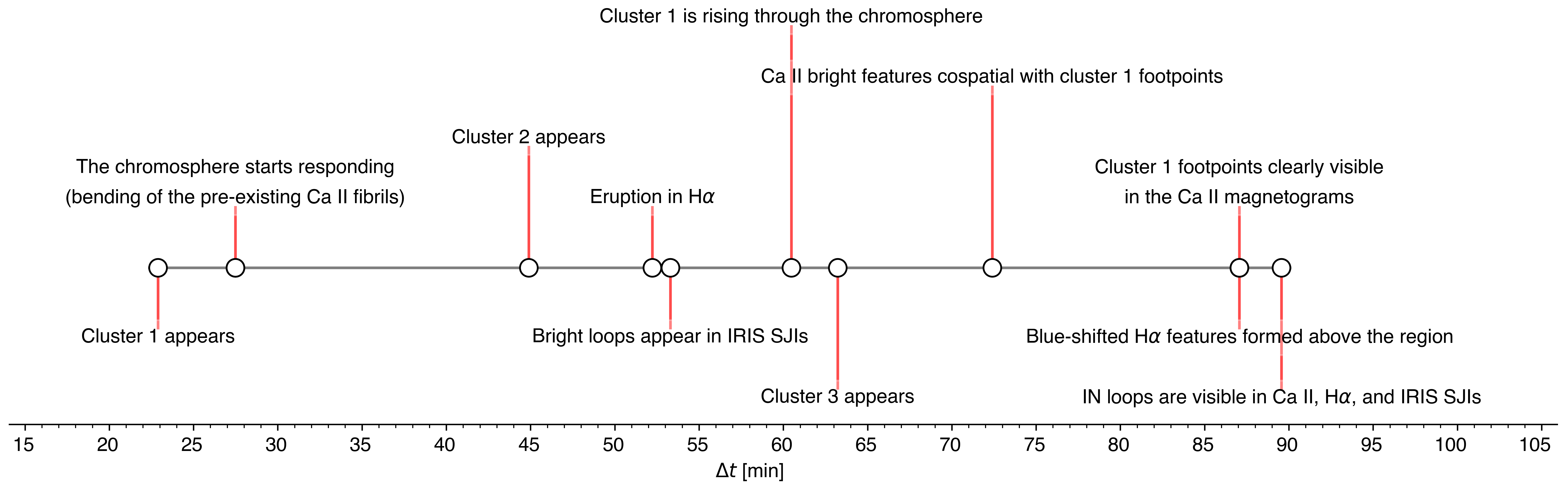}}
	\caption{Timeline for the flux emerging regions.}
	\label{timeline}
\end{figure*}

Finally, we note that the fields inferred from the \ion{Ca}{2} measurements may mix signals from different magnetic structures. The real chromospheric signal in cluster 1 starts to be visible from about $\Delta t$=70~min, in cluster 2 from $\Delta t$=80~min, and in cluster 3 from $\Delta t$=90~min. Before that, we may be confusing the emerging bipoles with the surrounding network canopy and photospheric signal. Once we start capturing more of the IN chromospheric signal, the fields become consistently weaker in the \ion{Ca}{2} line compared to the \ion{Fe}{1} line.

\subsection{Emergence of IN bipoles as seen by IRIS}

Before we describe the signatures of the emerging fields in the IRIS measurements, we would like to remind the reader that the SJI 1400 filter is sensitive to emission from the transition region \ion{Si}{4} 1394 and 1403 \AA\/ lines and the continuum formed in the upper photosphere/lower chromosphere. The latter source likely dominates in the bright features within NW elements that are visible in the IRIS SJI 2796 and SJI 1400 filtergrams from the start to the end of the observations. The bright features are connected by many rapidly changing loop-like structures (Figure \ref{iris_movie}). These loop-like features are likely caused by \ion{Si}{4} contributions to the passband. Except for the NW elements, the slit-jaw images do not show any stable signal co-spatial with the emerging magnetic fields. Starting from $\Delta t$=53:19, a bright feature appears at (5\arcsec, 6\arcsec). It is located at the same position as the negative footpoint in cluster 1. Similarly to NW features, this brightening most likely originates from the upper photosphere/lower chromosphere continuum. The reason is that the structure becomes visible at the same time as in the SJI 2796 and \ion{Ca}{2} blue wing intensity maps. From $\Delta t$=89:35, the emerging fields are in the chromosphere and the SJI 1400 images reveal a lot of bright loops between the footpoints, and also between the footpoints and NW patches. Many of these irregular bright loops are probably the result of the heating generated by reconnection of the rising and ambient magnetic field lines, and are likely caused by \ion{Si}{4} emission. This has been confirmed by studying the corresponding \ion{Si}{4} spectra (see further discussion below). 

The spectral shapes of the \ion{Mg}{2} h and k lines provide valuable information on the evolution of the magnetic field lines. The profiles shown in the lower right panel in Figure \ref{iris_profiles} indicate upflowing motions (according to \citealt{Leenaartsetal2013}, k2r/h2r peaks stronger than k2v/h2v peaks imply upflows in the upper chromosphere), suggesting that the fields are rising up, crossing the formation heights of the \ion{Mg}{2} h and k lines (also visible in Figure \ref{comparison}, which shows the velocity maps obtained with IRIS$^{2}$). 

To estimate the effects of the emerging IN magnetic fields on the thermal and dynamical properties of the upper solar atmosphere, we inverted the observed \ion{Mg}{2} spectral profiles using IRIS$^{2}$. The inversions reveal the highest temperatures at the locations of NW magnetic elements and above the emerging flux region, marked with the white box (Figure \ref{mgmapold}). The selected area is magnified in Figure \ref{comparison}. The temperature distribution above the emerging IN clusters at $\text{log}_{10}\tau_{500}=-5.8$ can be seen in the left panel. As the slit crosses the region it scans various loops between the footpoints. The temperature is higher along the loops, compared to the non-emerging QS flux regions. However, since the rasters were taken without compensating for solar rotation, the slit does not observe individual loops all the time, rather it provides spectra for various intermittent loops across the emerging regions. Our observations thus support a scenario in which the loops rise through the chromosphere and reconnect with the pre-existing overlying NW loops, and the associated currents cause heating and irregular brightenings along the field lines connecting the footpoints. Reconnection seems plausible since we find evidence for bidirectional reconnection flows in the FUV IRIS spectra. LOS plasma velocities within the emerging region are shown in the right panel of Figure \ref{comparison} and reveal that there are strong chromospheric upflows close to the loop tops, reaching speeds of about $-5$~km~s$^{-1}$, and downflows of $\sim10$~km~s$^{-1}$ at the positions of the footpoints. These results further establish our picture of plasma being pushed up to the chromosphere with plasma draining from the loops in the more vertically oriented footpoints. The temperature along the loops ranges from about 8 to 9~kK, and may even reach 10~kK, while inside the footpoints it can take values of up to 12~kK. This is about 2 to 6~kK hotter than the unperturbed atmosphere. 

Observations in the far-UV IRIS lines support a scenario in which these fields eventually reach the transition region where they reconnect with the ambient fields and produce bidirectional outflows. An example of reconnection events in the upper chromosphere/transition region can be seen around $\Delta t$=112, when the slit crosses the emerging fields and captures bright loops between the negative (north) and positive (south) polarity footpoints. Note that the solar rotation compensation was switched off during the observations, meaning that the IRIS slit was not observing the flux emerging region all the time, but was moving across the QS area.

A much more pronounced event happens around $\Delta t$=131:45, and is sufficiently strong to produce emission in the IRIS \ion{C}{2} 1335, \ion{Si}{4} 1394, and \ion{Si}{4} 1403 lines. We show in Figure \ref{iris_profiles} several UV spectral profiles obtained by IRIS at different locations along the slit (marked with the black, red, and violet horizontal lines in Figure \ref{iris_movie}), during the maximum emission. The transition region profiles shown here exhibit enhanced emission with broad wings of more than 100~km~s$^{-1}$. They are non-Gaussian, mostly triangular-shaped profiles similar to previous observations of plasmoid-mediated magnetic reconnection \citep{Innesetal1997, Innesetal2015, RouppevanderVoortetal2017}. Broad, multi-component UV profiles at reconnection sites were observed as well by \citeauthor{Guglielminoetal2018} (\citeyear{Guglielminoetal2018}; see also \citeauthor{Guoetal2020} \citeyear{Guoetal2020}) who associated them to small plasmoids moving at different velocities. The \ion{C}{2} and \ion{Si}{4} lines indicate that the Doppler velocities change from blueshifts to redshifts when moving from the southern to the northern parts of the emission enhancement. This is compatible with what would be expected from bidirectional outflows produced by reconnection events \citep[e.g.,][]{Innesetal1997, Peteretal2014}, which lead to greatly broadened spectral lines \citep{Innesetal2015}. In our observations, most of the IRIS FUV profiles during enhanced emission events show broad, non-Gaussian, blue- or red-shifted profiles. However, some of them are clearly double-peaked (Figure \ref{iris_profiles2}). 

\section{Discussion and Conclusions}
\label{sect5}

In this paper we studied the emergence of IN flux clusters and observed their impact on the QS chromosphere and transition region. We analyzed simultaneous SST and IRIS observations, sampling the photosphere, chromosphere, and transition region. Using SST magnetograms obtained from measurements in the \ion{Fe}{1} 6173 and Mg 5173~\AA\ lines, we identified and tracked all IN flux features that appear in three clusters emerging close to each other. Considering their locations, it is possible that all three clusters belong to the same, larger magnetic system. Thanks to our multi-wavelength observations, we were able to observe the entire temporal evolution of these magnetic clusters and their coupling to the upper solar atmosphere.

From SST \ion{Fe}{1} 6173 magnetograms we determined that the total flux brought to the solar surface by the clusters is $1.9\times10^{18}$, $2.5\times10^{18}$, and $5.3\times10^{18}$~Mx. None of the individual flux patches surpasses the $3\times10^{18}$ threshold, which suggest that with lower resolution, many of the detected patches would be invisible and the clusters would probably not be classified as ephemeral regions. Indeed, only parts of the largest cluster were seen in HMI (and would barely be visible to MDI). Had the analysis been based only on such instruments, the interpretation and understanding of the emerging flux would have been more challenging. However, the morphology of the emerging fields resembles that of ephemeral regions. Therefore, the fields analyzed here support the idea of a continuous flux and size distribution of magnetic elements in the solar IN \citep{Gosic2015}.

In the photosphere, the emergence of magnetic fields reveals the pattern expected from earlier observations and theoretical works \citep{Centenoetal2007, Cheungetal2007, MartinezSykoraetal2008, TortosaAndreuMorenoInsertis2009, MartinezGonzalezBellotRubio2009, MartinezGonzalezetal2010, Guglielminoetal2008, Guglielminoetal2010, Kontogiannisetal2019}. It is worth noting that thanks to the high resolution and sensitivity of the SST observations, the flux emergence events presented in this paper show a much higher complexity. Emerging magnetic fields are first visible as patches of linear polarization signal, and later as circular polarization patches that appear at the edges of granular cells and move towards the intergranular lanes. We also detect elongated granules with maximum upflows of about $-1$~km~s$^{-1}$. These values are in agreement with observational results reported by \cite{Guglielminoetal2008} and \cite{Guglielminoetal2012}, and also with the theoretical estimates by \cite{Cheungetal2008}. Magnetic elements within the clusters appeared as mixed-polarity features that often merge with like-polarity features and cancel with opposite-polarity patches. In general, the clusters share the same magnetic axis and form a large-scale magnetic bipole with clearly distinguishable positive and negative polarity footpoints in the SST \ion{Fe}{1} and \ion{Mg}{1} b$_{2}$ magnetograms. 

One of the most important results of this paper is the discovery that IN magnetic fields with strengths between ~450 and 800~G can reach the chromosphere. This has been demonstrated from full spectropolarimetric measurements in the \ion{Ca}{2} 8542~\AA\ line. These magnetograms show the strongest footpoints that were created through merging of the same polarity magnetic elements from the three clusters. 

Figure \ref{timeline} summarizes the temporal evolution of the three IN clusters and the atmospheric response to them. According to our observations, it took about one hour for the IN fields to fully break through the ambient fields and emerge in the chromospheric layers. This is very different from the rising times reported by \cite{MartinezGonzalezBellotRubio2009}, and should be further investigated. It is possible that the pace at which magnetic fields rise through the solar atmosphere is related to different preexisting magnetic configurations. If such a delay is typical, it may be quite difficult to connect heating events in the chromosphere to individual flux emergence events without the type of extensive datasets we have used here, which show the full evolution from the photosphere to the transition region for long periods of time. Moreover, these results set important physical constraints to MHD models of the QS and IN regions, i.e., we should regularly see IN fields in the chromosphere, at least the strongest ones. Therefore, the question should not be whether IN bipoles can reach the upper atmosphere, but rather what fraction of them do so. It is important to continue investigating, both observationally and theoretically, how convective upflows and/or magnetic buoyancy drive the emergence of small-scale fields and what the role of the ambient field is. It is expected that state-of-the-art radiative-magnetohydrodynamical simulations of the QS as in \cite{MartinezSykoraetal2019} and future measurements with the Daniel K. Inouye Solar Telescope \citep[DKIST;][]{Elmoreetal2014} and the European Solar Telescope (EST; \citeauthor{Colladosetal2013} \citeyear{Colladosetal2013}; see also \citeauthor{Schlichenmaieretal2019} \citeyear{Schlichenmaieretal2019}) will allow us to better understand this topic. The next generation telescopes should be capable of detecting IN fields weaker than those observed until now, which will hopefully be sufficient to consistently detect circular and linear polarization signals even in chromospheric lines.

During their lifetimes, IN magnetic fields seem to constantly perturb the overlying atmosphere. We detect absorption chromospheric features that can be understood as the acceleration of chromospheric material driven by strong thermal and magnetic pressure gradients during the expansion of the emerging loops. This may lead to the generation of cool jets and subsequent shocks in the overlying atmospheric layers \cite{NobregaSiverioetal2016, NobregaSiverioetal2017}. In addition to this, IRIS observations reveal another important piece of information---small-scale IN loops can reach even the transition region. SJI 1400 filtergrams show bright loops connecting the footpoints, while the IRIS FUV spectral lines reveal clear, strong emission produced by the emergence of IN fields. The line profiles indicate reconnection processes happening in the upper chromosphere and the transition region as a result of reconnection between pre-existing ambient fields and the IN fields emerging into the chromosphere. We estimated from the IRIS$^{2}$ inversions of the \ion{Mg}{2} h and k lines that the chromospheric temperature above the emerging IN fields rises up to 10~kK (from 5-6~kK in the very quiet regions). The inferred chromospheric upflow (due to plasma being pushed up to the chromosphere) and downflow velocities (caused by plasma draining from the loop apex to the footpoints) are between $5$ and $10$~km~s$^{-1}$, which is similar to the values reported for small-scale emerging loops in ARs \citep{Ortizetal2014}. 

From our SST and IRIS observations, it is clear that IN loops can reach chromospheric and transition region heights, and locally heat the upper solar atmosphere. To make further advances in our knowledge of small-scale flux emergence, we plan to carry out a statistical study to determine how many of the newly appeared IN fields reach the chromosphere, and how they affect globally the energetics and dynamics of the upper atmosphere in the QS. We will also study the properties of IN magnetic loops and the solar atmospheres above them in order to understand what determines whether IN loops are able to break through the chromosphere and reach the atmospheric layers above.

\acknowledgments 
IRIS is a NASA Small Explorer Mission developed and operated by LMSAL with mission operations executed at NASA Ames Research Center and major contributions to downlink communications funded by ESA and the Norwegian Space Centre. MG, BDP, and ASD are supported by NASA contract NNG09FA40C (IRIS). The work of LBR and SEP was supported by the Spanish Ministerio de Econom\'ia and Competitividad through grants ESP2013-47349-C6-1-R and ESP2016-77548-C5-1-R, including a percentage from European FEDER funds. LBR acknowledges financial support from the State Agency for Research of the Spanish Ministerio de Ciencia e Innovaci\'{o}n through project RTI2018-096886-B-C5 (including FEDER funds) and through the “Center of Excellence Severo Ochoa" award to the Instituto de Astrof\'isica de Andaluc\'ia (SEV-2017-0709). The Swedish 1 m Solar Telescope is operated on the island of La Palma by the Institute for Solar Physics of Stockholm University in the Spanish Observatory del Roque de los Muchachos of the Instituto de Astrof\'isica de Canarias. The Institute for Solar Physics is supported by a grant for research infrastructures of national importance from the Swedish Research Council (registration number 2017-00625). This research has made use of NASA’s Astrophysics Data System.

\end{document}